\documentclass{nature}
\usepackage{graphicx,psfrag}
\usepackage{mathrsfs}
\usepackage{amsmath,amsfonts,amssymb}
\usepackage{multirow}
\usepackage{comment}
\usepackage{units}
\usepackage{url}
\usepackage{enumerate}
\usepackage[normalem]{ulem} 
\usepackage{longtable}
\usepackage{threeparttablex}
\usepackage[rgb]{xcolor}
\usepackage{gensymb}
\usepackage{adjustbox}
\usepackage{ccaption}
\usepackage{subcaption}
\usepackage{aas_macros}
\usepackage[none]{hyphenat}
\usepackage{etoolbox}
\makeatletter
\renewcommand{\citen}[1]{%
    \begingroup
    \let\@cite\@firstofone
    \cite{#1}%
    \endgroup
}
\makeatother

\newcommand\arcsec{\mbox{$^{\prime\prime}$}}% 

\makeatletter
\newcommand{\dummylabel}[2]{\def\@currentlabel{#2}\label{#1}}
\makeatother

\title{Millihertz Oscillations Near the Innermost Orbit of a Supermassive Black Hole} % this is 75 characters

\author{Megan Masterson$^{1,*}$, Erin Kara$^{1}$, Christos Panagiotou$^{1}$, William N. Alston$^{2}$, Joheen Chakraborty$^{1}$, Kevin Burdge$^{1}$, Claudio Ricci$^{3,4}$, Sibasish Laha$^{5,6,7}$, Iair Arcavi$^{8}$, Riccardo Arcodia$^{1}$, S. Bradley Cenko$^{5,9}$, Andrew C. Fabian$^{10}$, Javier A. Garc\'{i}a$^{11,12}$, Margherita Giustini$^{13}$, Adam Ingram$^{14}$, Peter Kosec$^{15}$, Michael Loewenstein$^{5,16}$, Eileen T. Meyer$^{17}$, Giovanni Miniutti$^{13}$, Ciro Pinto$^{18}$, Ronald A. Remillard$^{1}$, Dev R. Sadaula$^{5,6,7}$, Onic I. Shuvo$^{17}$, Benny Trakhtenbrot$^{8}$, Jingyi Wang$^{1}$} 

\begin{document}

\maketitle

\begin{affiliations}
 \item MIT Kavli Institute for Astrophysics and Space Research, Massachusetts Institute of Technology, Cambridge, MA 02139, USA
 \item Centre for Astrophysics Research, Department of Physics, Astronomy and Mathematics, University of Hertfordshire, College Lane, Hatfield, AL10 9AB, UK
 \item N\'{u}cleo de Astronom\'{i}a de la Facultad de Ingenier\'{i}a, Universidad Diego Portales, Av. Ej\'{e}rcito Libertador 441, Santiago, Chile
 \item Kavli Institute for Astronomy and Astrophysics, Peking University, Beijing 100871, People's Republic of China
 \item Astrophysics Science Division, NASA Goddard Space Flight Center, 8800 Greenbelt Road, Greenbelt, MD 20771, USA
 \item Center for Space Science and Technology, University of Maryland Baltimore County, 1000 Hilltop Circle, Baltimore, MD 21250, USA
 \item Center for Research and Exploration in Space Science and Technology, NASA/GSFC, Greenbelt, Maryland 20771, USA
 \item School of Physics and Astronomy, Tel Aviv University, Tel Aviv 69978, Israel
 \item Joint Space-Science Institute, University of Maryland, College Park, MD 20742, USA
 \item Institute of Astronomy, University of Cambridge, Madingley Road, Cambridge CB3 0HA, UK
 \item X-ray Astrophysics Laboratory, NASA Goddard Space Flight Center, Greenbelt, MD 20771, USA
 \item Cahill Center for Astronomy and Astrophysics, California Institute of Technology, Pasadena, CA 91125, USA
 \item Centro de Astrobiolog\'{i}a (CAB), CSIC-INTA, Camino Bajo del Castillo s/n, Villanueva de la Cañada, 28692 Madrid, Spain
 \item School of Mathematics, Statistics, and Physics, Newcastle University, Newcastle upon Tyne NE1 7RU, UK
 \item Center for Astrophysics $\mid$ Harvard \& Smithsonian, Cambridge, MA 02138, USA
 \item Department of Astronomy, University of Maryland, College Park, MD 20742, USA
 \item Department of Physics, University of Maryland Baltimore County, 1000 Hilltop Circle Baltimore, MD 21250, USA
 \item INAF - IASF Palermo, Via U. La Malfa 153, I-90146 Palermo, Italy
\end{affiliations}

% \begin{bibunit}[naturemag]

\begin{abstract}

Recent discoveries from time-domain surveys are defying our expectations for how matter accretes onto supermassive black holes (SMBHs). The increased rate of short-timescale, repetitive events around SMBHs, including the newly-discovered quasi-periodic eruptions (QPEs)\cite{Miniutti2019,Giustini2020,Arcodia2021,Chakraborty2021,Arcodia2024a}, are garnering further interest in stellar-mass companions around SMBHs and the progenitors to mHz frequency gravitational wave events. Here we report the discovery of a highly significant mHz Quasi-Periodic Oscillation (QPO) in an actively accreting SMBH, 1ES\,1927+654, which underwent a major optical, UV, and X-ray outburst beginning in 2018\cite{Trakhtenbrot2019,Ricci2020}. The QPO was first detected in 2022 with a roughly 18-minute period, corresponding to coherent motion on scales of less than 10 gravitational radii, much closer to the SMBH than typical QPEs. The period decreased to 7.1 minutes over two years with a decelerating period evolution ($\ddot{P} > 0$). This evolution has never been seen in SMBH QPOs or high-frequency QPOs in stellar mass black holes. Models invoking orbital decay of a stellar-mass companion struggle to explain the period evolution without stable mass transfer to offset angular momentum losses, while the lack of a direct analog to stellar mass black hole QPOs means that many instability models cannot explain all of the observed properties of the QPO in 1ES\,1927+654. Future X-ray monitoring will test these models, and if it is a stellar-mass orbiter, the Laser Interferometer Space Antenna (LISA) should detect its low-frequency gravitational wave emission.

\end{abstract}

\begin{figure}
    \centering
    \includegraphics[width=\textwidth]{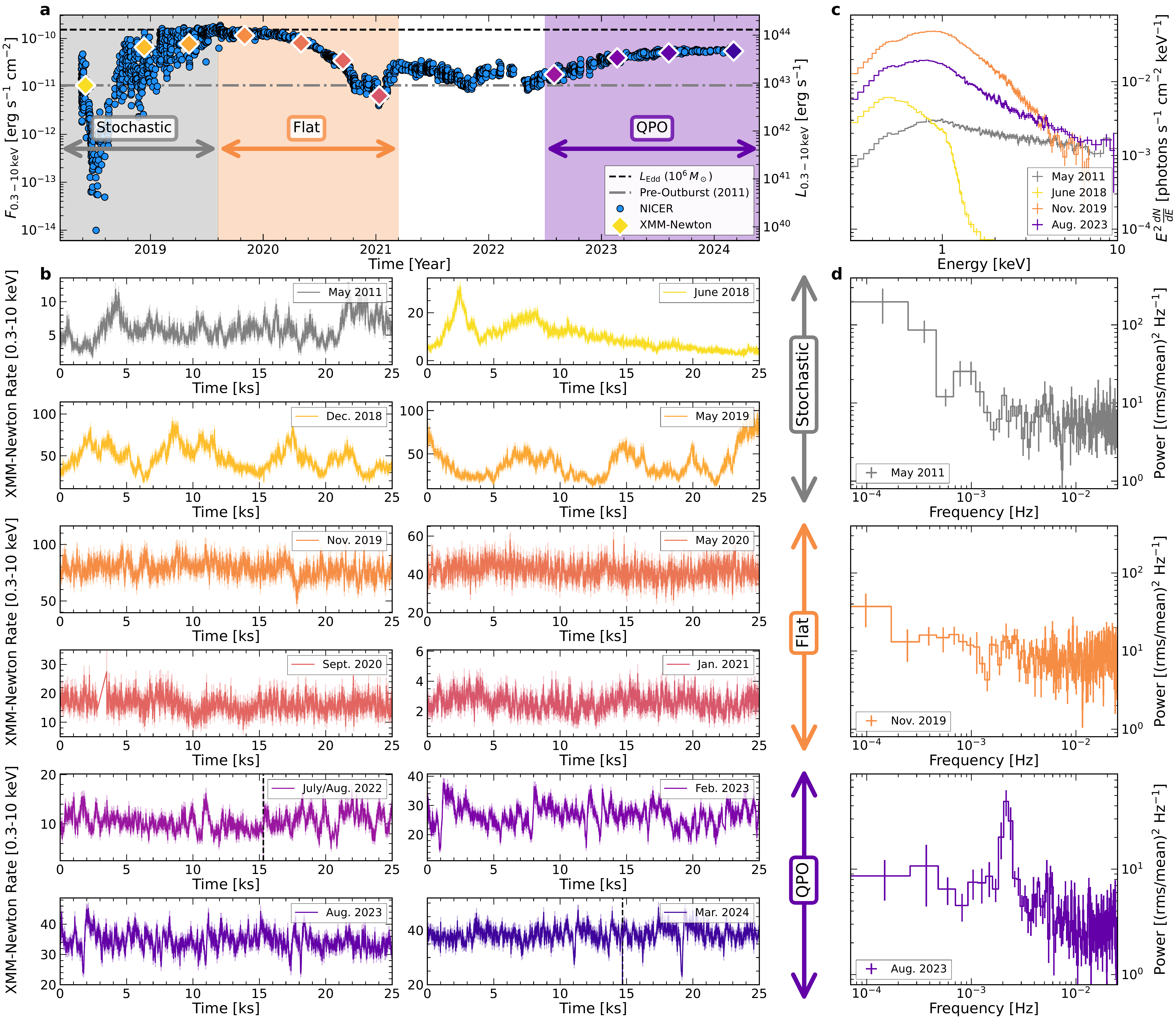}
    \caption{\linespread{1}\small X-ray spectral-timing overview of 1ES\,1927+654. \textbf{a,} Long-term X-ray light curve (0.3-10 keV) beginning 2 months after the optical outburst. NICER (XMM-Newton) data is shown with blue circles (diamonds, colored by time). The observed flux and luminosity have been scaled from count rates assuming a $\Gamma = 3$ power-law spectrum. The horizontal black dashed line shows the Eddington limit for a $10^6\, M_\odot$, and the grey dot-dashed line shows the pre-outburst flux level from an XMM-Newton observation in 2011\cite{Gallo2013}. Three different variability classes, stochastic, flat, and QPO, are shaded in grey, orange, and purple, respectively. \textbf{b,} XMM-Newton light curves (0.3-10 keV) with 20s time binning for each of the points shown in the top panel. All panels only show a 25~ks segment, which allows for a direct comparison of variability timescales. Both the July-August 2022 and March 2024 panels shows light curves from two observations taken within roughly one week, and the vertical dashed line separates them. \textbf{c,} Subset of the XMM-Newton spectra, highlighting the dramatic evolution in spectral shape. Of these four observations, the QPO is only present in the August 2023 data shown in purple. The yellow spectrum from June 2018 highlights the destruction of the X-ray corona that occurred early in the optical outburst of 1ES\,1927+654. \textbf{d,} 2-10 keV PSDs for three of the observations shown to the left (binned to $n = 6$ frequencies per bin with standard error uncertainties). The short timescale variability of 1ES\,1927+654 has evolved drastically over the last six years and clearly shows the build up of a rapid quasi-periodicity.}
    \label{fig:overview}
\end{figure}

1ES\,1927+654 ($z = 0.019422$, $M_\mathrm{BH} = 1.38_{-0.66}^{+1.25} \times 10^6 \, M_\odot$; ref. \citenum{Li2022}) underwent a major outburst that was first detected by ASAS-SN in March 2018. Less than a month after the initial detection, this previously ``bare Seyfert 2"\cite{Boller2003,Tran2011,Gallo2013} (i.e. an AGN lacking broad emission lines, but with no X-ray obscuration) showed the formation of broad Balmer lines, becoming the first AGN caught changing between different spectral types in real time\cite{Trakhtenbrot2019}. X-ray observations beginning in May 2018 revealed an extremely soft, thermal spectrum ($kT_\mathrm{bb} \approx 80$ eV), indicating that the canonical X-ray corona, which produces the hard, power-law spectrum and had previously been observed in this system\cite{Gallo2013}, had been destroyed\cite{Ricci2020}. The X-ray flux (0.3-10 keV) then dropped by a factor of 1,000 over the span of two months, before undergoing a significant resurgence, with order of magnitude variations in flux on timescales of less than a day. Ultimately, the X-ray flux plateaued at $L_X \approx 10^{44}$ erg s$^{-1}$ (near the Eddington limit for a $10^6\, M_\odot$ black hole), followed by a year-long smooth decline marked by the strengthening of the X-ray corona\cite{Ricci2021,Masterson2022} (see Figure \ref{fig:overview}). One potential explanation proposed for this dramatic variability, including the destruction and recreation of the canonical X-ray corona, is that a tidal disruption event (TDE) occurred in the existing AGN\cite{Ricci2020}, thereby shocking and rapidly depleting the inner accretion flow. 

\begin{figure}
    \centering
    \includegraphics[width=\textwidth]{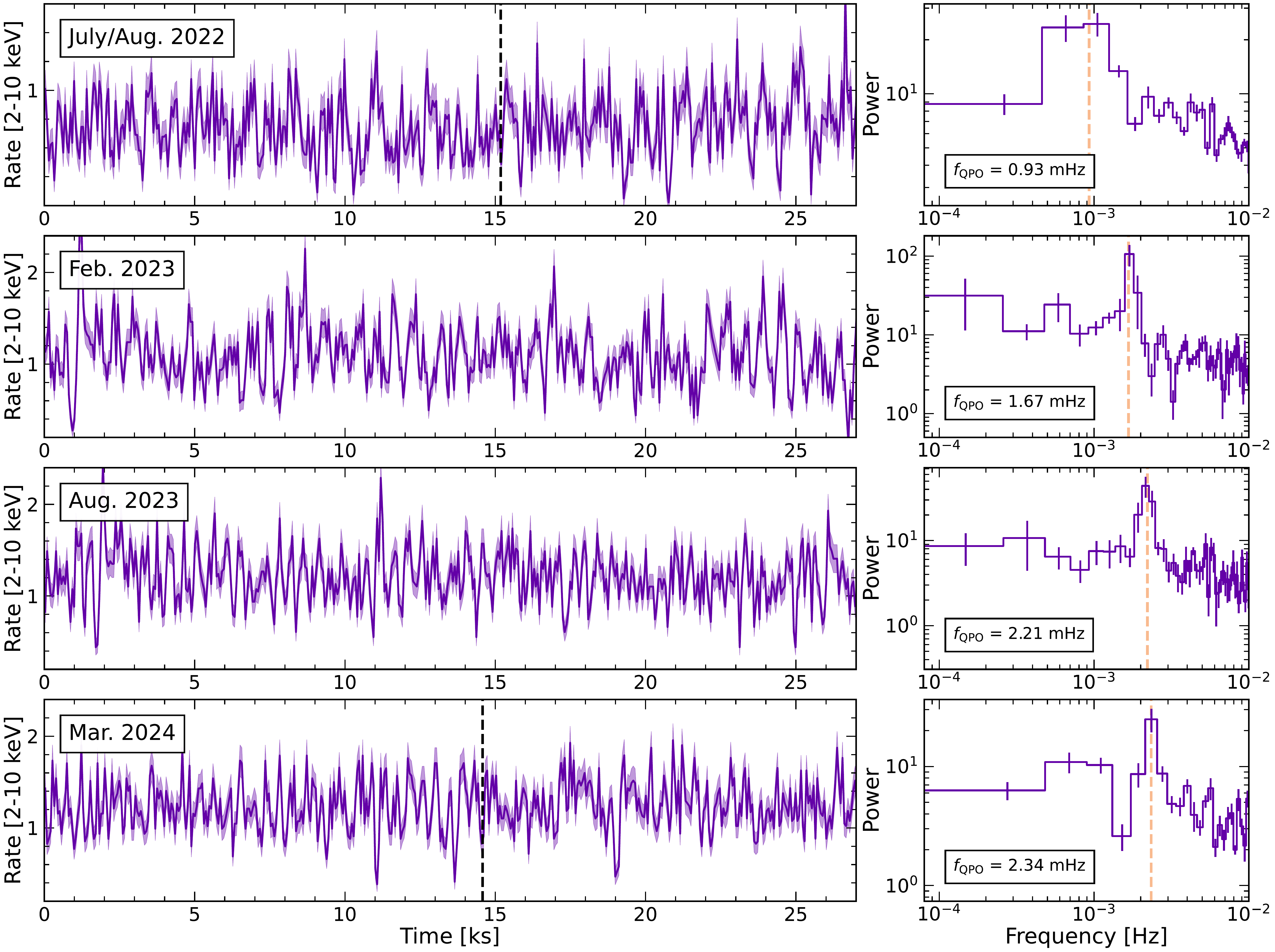}
    \caption{\linespread{1}\normalsize XMM-Newton light curves (left) and PSDs (right) in the 2-10 keV band for the observations which show a QPO. The shaded regions in the light curves represent the Poisson uncertainty. To highlight the significant evolution of the QPO frequency, the $x$-axes of both the light curves and the PSDs are fixed to the same range. The light curves are binned to 60s to highlight the QPO by eye, while the PSDs are made with 20s binned light curves to increase the frequency resolution. Roughly 50 QPO cycles are observable in each of these short observations with XMM-Newton. The top and bottom panels show data from 2 observations taken within roughly 1 week of one another due to limited visibility with XMM-Newton. The vertical dashed lines show where the first observation stops and the next begins. For these observations, we computed individual PSDs from each observation and then binned them together by averaging at each frequency to make the PSDs shown on the right. All PSDs are then averaged over neighboring frequencies with $n = 6$ frequencies per bin and show standard error uncertainties.}
    \label{fig:lc_psd}
\end{figure}

By early 2021, 1ES\,1927+654 had returned to its pre-outburst X-ray flux and spectroscopic state\cite{Masterson2022,Laha2022}, suggesting a potential end to the dramatic outburst that began in 2018. However, in early 2022 the X-ray flux began to rise again, this time with no coincident optical or UV flare detected\cite{Ghosh2023}. During this rise in the X-ray flux, the corona remained intact, albeit at a relatively soft level ($\Gamma \approx 3$) compared to typical AGN ($\Gamma \approx 1.6-2$)\cite{Liu2016}. High-cadence NICER observations during this rise do not show the dramatic order of magnitude variability seen in the 2018 dip and subsequent flare. However, observations with XMM-Newton from July 2022 to March 2024 reveal quasi-periodic variability at the $\approx 10\%$ level on timescales of $\approx 400-1000$s, associated with a mHz frequency QPO (see Figures \ref{fig:overview} and \ref{fig:lc_psd}). QPOs are notoriously difficult to detect in SMBHs, due in part to their relatively long timescales compared to a typical observation duration\cite{Vaughan2005a}; this also explains why we do not detect the QPO with NICER, as the QPO period is roughly the same duration as a single NICER snapshot. To date, only a handful of accreting SMBHs show similarly high-significance mHz QPOs, including the super-Eddington narrow-line Seyfert 1 AGN RE J1034\cite{Gierlinski2008,Alston2014}, the canonical TDE ASASSN-14li\cite{Pasham2019}, and the TDE/QPE candidate 2XMM J1231\cite{Lin2013,Lin2017,Webbe2023}.

Figure \ref{fig:lc_psd} shows the XMM-Newton light curves and power spectral densities (PSDs; see Methods, Section \ref{sec:timing}) in the 2-10 keV band for each of the four epochs in which the QPO is detected. The apparent periodicity is evident in both the light curves and the PSDs, which show a narrow excess (the QPO) on top of broadband noise. To characterize the QPO, we first fit the data with two broadband noise models, a power-law and a Lorentzian centered at $f = 0$. We assessed the statistical significance of the QPO using two methods (see Methods, Section \ref{subsec:significance} for details): (1) we simulated $10^5$ light curves based on the two best-fit broadband noise models\cite{Timmer1995}, computed their PSDs, and tested how often we recovered a peak as significant as the one we observed, and (2) we compared the Akaike Information Criterion (AIC)\cite{Akaike1974} between the PSD fit with broadband noise only versus fit with broadband noise and an additional Lorentzian for the QPO. Both methods point to highly significant QPO features ($p \lesssim 0.01$).

The first detection of the QPO occurred in July-August 2022, in four short ($\approx$ 15 ks) observations taken within one week of one another (due to limited visibility). The QPO was not significantly detected in any observation individually, but a similar excess shows up at roughly 0.9 mHz in all four observations, suggesting a weak and broad QPO (see Methods, Section \ref{subsec:significance} for details). By February 2023, the QPO was stronger, narrower, and at a significantly higher frequency of $1.67 \pm 0.04$ mHz. This detection is highly significant (at $\approx 6.5\sigma$, from the AIC method) and was possible with only a single short ($\approx 30$ ks) observation with XMM-Newton. Continued monitoring with roughly 6 month cadence showed that QPO persisted for another year and the frequency continued to evolve. In August 2023, the QPO was detected at $2.21 \pm 0.05$ mHz (at $\approx 6.8\sigma$), and in March 2024, it was detected (in one of two observations) at only marginally higher frequency of $2.34 \pm 0.05$ mHz (at $\approx 5.2\sigma$). These three observations show a total of 45, 59, and 34 cycles, respectively, adding to the confidence in these features.

\begin{figure}
    \centering
    \includegraphics[width=\textwidth]{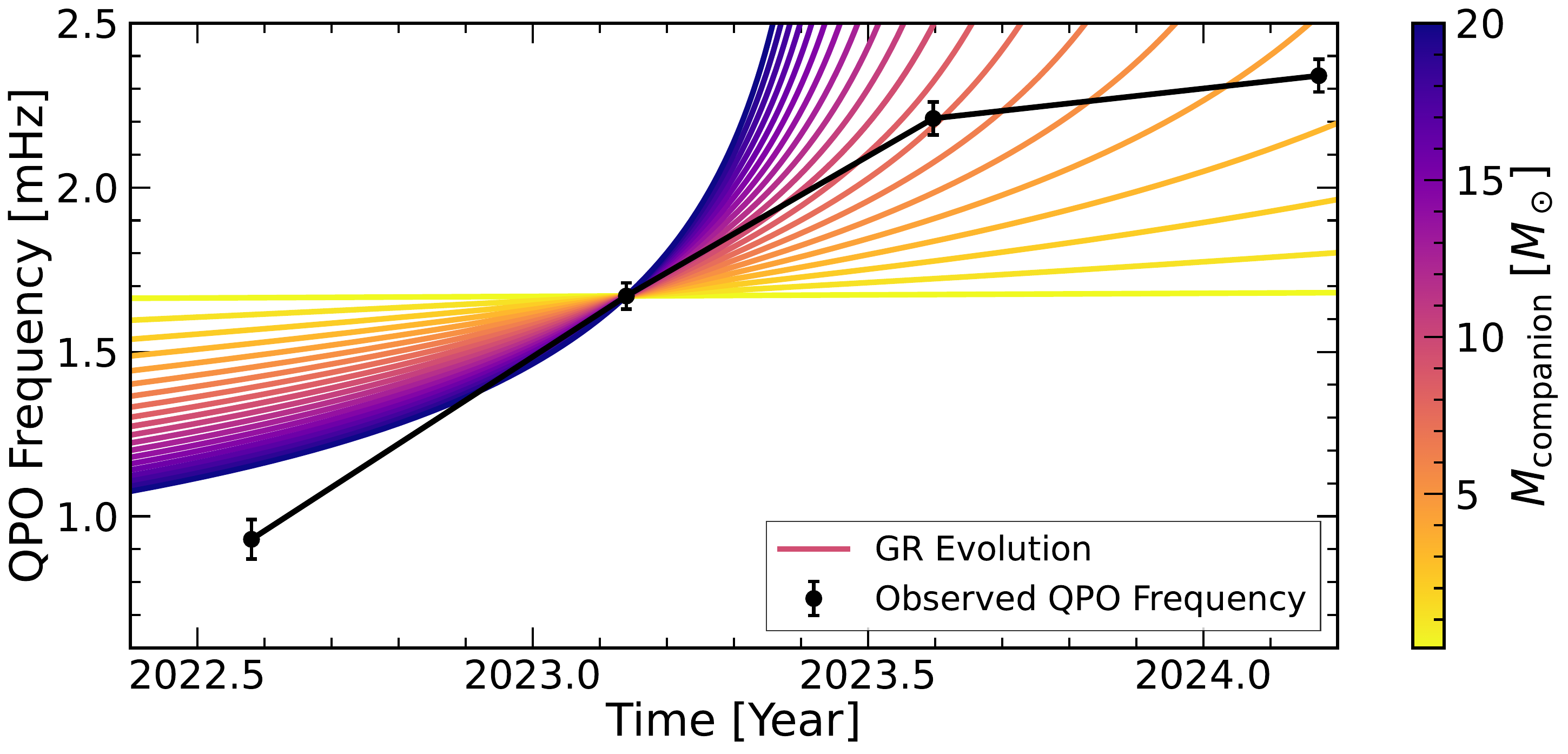}
    \caption{\linespread{1}\normalsize Evolution of the QPO frequency over time. The black points show the observed QPO frequency and 1$\sigma$ error bars, obtained by fitting an additional Lorentzian for the QPO. The colored lines show the expected evolution of an extreme mass ratio companion under GR alone, assuming a circular orbit ($e = 0$) and companion masses ranging from 0.1-20 $M_\odot$. The color bar shows the mass distribution. The GR model was chosen to match the QPO frequency in February 2023, as this was the data with the most significant detection and the lowest uncertainty on the QPO frequency. GR alone from an orbiting companion cannot account for the frequency evolution seen in 1ES\,1927+654.}
    \label{fig:qpo_freq_evol}
\end{figure}

One of the most peculiar features about the QPO in 1ES\,1927+654 is its significant frequency evolution with time (Figure \ref{fig:qpo_freq_evol}). The QPO frequency, $f$, is increasing ($\dot{f} > 0$), but shows a clear negative $\ddot{f}$; that is, the rate of increase of the QPO frequency is slowing down with time. This is unprecedented among the few QPOs that have been seen around SMBHs. RE J1034, the first AGN with a significant QPO detected in the mHz regime, shows only a roughly 10\% change to the QPO frequency that does not follow a similar monotonic evolution\cite{Alston2014,Xia2024}. ASASSN-14li, a TDE with a persistent 7.65 mHz QPO, showed a remarkably stable QPO frequency over 450 days despite a factor of 4 drop in the X-ray flux\cite{Pasham2019}. 

In addition to the frequency evolution, there are a few additional constraining observables of the QPO in 1ES\,1927+654 that can help inform the driving mechanism. First, the QPO is most prevalent in hard X-rays; the fractional root mean square (RMS) of the QPO increases from $\lesssim 7$\% in 0.3-2 keV, up to $\approx 15-20$\% in 2-10 keV, consistently across all observations (see Extended Data Figures \ref{fig:PSD_energy} and \ref{fig:rms}). This suggests that the QPO is primarily associated with the X-ray corona. Additionally, the QPO is highly coherent, with a typical $Q$ factor of 10 (where $Q = f_\mathrm{QPO}$/FWHM$_\mathrm{QPO}$), similar to what has been seen in QPOs around both SMBHs\cite{Gierlinski2008,Alston2014,Pasham2019} and black hole X-ray binaries (BHXBs)\cite{Remillard2006a,Ingram2019}. Lastly, the QPO frequency correlates strongly with the X-ray flux and the power-law spectral shape (see Extended Data Figure \ref{fig:qpofreq_fluxevol}).

1ES\,1927+654 shows suggestive similarities to the recently discovered class of nuclear transients called quasi-periodic eruptions (QPEs\cite{Miniutti2019,Giustini2020,Arcodia2021,Chakraborty2021,Arcodia2024a}; see Methods, Section \ref{subsec:qpes}), which show repetitive soft X-ray flares on timescales of hours to days. Like QPEs, 1ES\,1927+654 showed order of magnitude X-ray flux variations on timescales of hours to days in 2018-2019\cite{Ricci2021,Masterson2022}, has a relatively soft X-ray spectrum compared to standard AGN, and hosts a low mass SMBH ($M_\mathrm{BH} \approx 10^6\, M_\odot$)\cite{Li2022,Wevers2022}. Interestingly, both QPEs (including QPE candidates; see Methods Section \ref{subsec:qpes}) and 1ES\,1927+654 show potential connections to TDEs\cite{Miniutti2019,Ricci2020,Chakraborty2021} and super-soft AGN that appear to intrinsically lack a broad line region\cite{Boller2003,Terashima2012,Lin2013,Miniutti2013,Sun2013}. One of the leading models for QPEs is a stellar-mass companion striking a newborn accretion disk (fed by a TDE) twice per orbit\cite{Franchini2023,Linial2023}. For the hours-to-days recurrence times for QPEs, this model requires companions on slightly eccentric orbits at roughly 40-400 $R_g$, where $R_g = GM/c^2$ (although more eccentric orbits are allowed if the QPEs are driven instead by circularization shocks\cite{Lu2023} or periodic mass transfer from a companion\cite{King2022}).

If the QPO in 1ES 1927+654 is produced by an extreme mass ratio companion, it must be much closer to the SMBH ($\lesssim 10\, R_g$) than most QPEs, given the shorter QPO period. At such a small radius, the timescale for inspiral due to the emission of gravitational waves (GWs) is significantly shorter than for QPEs and can even be comparable to the duration of our observing campaign. However, the observed frequency evolution does not match the expected ``chirp" from GWs where the frequency increases and accelerates closer to the SMBH (i.e. $\ddot{f} > 0$, see Figure \ref{fig:qpo_freq_evol}). Instead, the frequency evolution appears qualitatively similar to the expectation for the onset of stable mass transfer from a degenerate orbiting companion,  in which the additional angular momentum from mass transfer can counteract the angular momentum losses from GW emission and gas drag to reverse the orbital evolution and drive the companion outward\cite{Marsh2004}. In the case of 1ES\,1927+654, a low mass ($\approx 0.1\, M_\odot$) white dwarf (WD) could survive at $\lesssim 10\, R_g$ and fill its Roche lobe at the observed 7.1 minute period (see Methods Section \ref{subsec:EMR}). Connecting this companion to the TDE proposed to explain the initial outburst is possible, but it is difficult to get a remnant so close to black hole within four years of the initial disruption (see Methods Section \ref{subsec:EMR} for further discussion). Additionally, it is still unclear how the companion interaction would couple to the X-ray corona, which is where the QPO is the strongest. Despite these difficulties, the orbiter model makes a key observable prediction -- if there is a 0.1 $M_\odot$ WD in orbit around the SMBH, then, at 84 Mpc, it should be detectable with next generation mHz gravitational wave detectors (e.g. LISA). 

In addition to QPEs, we can also draw on the vast amount of work on QPOs in BHXBs, which are far more abundant, to understand the physics driving the QPO in 1ES\,1927+654. The QPO coherence, dominance in the hard X-rays, and fractional RMS are similar to both Type C low-frequency QPOs (LFQPOs) and high-frequency QPOs (HFQPOs) in BHXBs. Despite occurring primarily in the Comptonized component, HFQPOs only appear in the very high, steep power-law state\cite{Remillard2002,Remillard2006b}, which is potentially analogous to the current spectral state of 1ES\,1927+654  ($\Gamma \approx 3$ with a strong soft excess\cite{Ghosh2023}; see top right panel of Figure \ref{fig:overview}). However, HFQPOs tend to show relatively stable QPO frequencies\cite{Remillard2002,Remillard2006b}, unlike what we see in 1ES\,1927+654, although these features are rare and difficult to detect\cite{Belloni2012}. On the other hand, Type C LFQPOs occur in the hard and hard-intermediate states and regularly show a correlation between QPO frequency and the flux and spectral shape\cite{Vignarca2003}, as is seen in 1ES\,1927+654. However, scaling an $f \approx 10$ Hz QPO around $M_\mathrm{BH} \approx 10\, M_\odot$ down to mHz frequencies would imply $M_\mathrm{BH} \lesssim 5 \times 10^4 \, M_\odot$ for 1ES\,1927+654. During the original outburst, the luminosity in the X-ray band alone reached $L_X \approx 10^{44}$ erg s$^{-1}$, which, assuming $M_\mathrm{BH} \approx 5 \times 10^4 \, M_\odot$, yields an unphysically large Eddington ratio ($L \gtrsim 20\, L_\mathrm{Edd}$). Thus, there is no direct analog between the QPO in 1ES\,1927+654 and the QPOs in BHXBs, but the physics put forth to explain QPOs in BHXBs could still be similar.

Many models for BHXB QPOs invoke instabilities and geometric effects to explain QPO phenomenology. One of the most promising class of models invokes Lense-Thirring precession from the misalignment between the inner accretion flow and the black hole spin axis\cite{Stella1999,Ingram2009}. Solid-body precession of the inner hot flow can naturally reproduce both the high energy nature of the QPO (i.e. the coupling to the corona) and the correlation between the QPO frequency and the X-ray flux\cite{Ingram2009}. A truncated accretion disk and inner hot flow can be produced by disk tearing instabilities, which have been seen in simulations\cite{Nixon2012a,Liska2021,Musoke2023} (although these simulations often require extreme conditions such as high spin, significant misalignment, or small disk aspect ratios). However, for 1ES\,1927+654, even the extreme case of Lense-Thirring precession of a single test particle would require that the SMBH is near maximally spinning and the QPO arises from within a few gravitational radii, which is difficult to reconcile with the long-lived nature of the QPO. Recent GRMHD simulations have shown that disk tearing and precession can also give rise to HFQPOs from radial epicyclic modes near the disk tearing radius\cite{Musoke2023}. To reproduce the observed QPO frequency with the radial epicyclic frequency, the black hole must be rapidly spinning ($a \gtrsim 0.8$), the SMBH mass must be low ($M \lesssim 10^6 \, M_\odot$), and the disk tearing radius must be small ($\lesssim 5 R_g$; see Methods, Section \ref{subsec:instability}). Additionally, the HFQPOs in these simulations are not long-lived; when scaled up to SMBH masses, they tend to last for $\sim$ days, much shorter than the duration in which we detect QPOs in 1ES\,1927+654. However, these same GRMHD simulations also show long-term variability and depletion of the inner accretion flow, which was suggested to be a potential driver of changing-look AGN and akin to what was seen in the early stages of the 2018 outburst of 1ES\,1927+654\cite{Kaaz2023}. Additionally, misalignment between the inner disk and black hole spin axis can be naturally explained in the TDE picture for the initial outburst\cite{Ricci2020}, but it is then unclear why the QPO appeared only at late times ($\approx$ 4 years after the outburst began). 

Finally, the high energy of the QPO, coupled with the fact that the QPO turns on around the same time that a newly launched jet is observed (through a rapid rise in radio flux; see ref. \citenum{Meyer2024}), suggests a potential connection to the intrinsic properties of the corona/jet. Thus, we consider an additional model for LFQPOs in BHXBs that may lend itself well to 1ES\,1927+654 -- resonant oscillations in the corona driven by the propagation of magnetoacoustic waves\cite{Cabanac2010,Buisson2019}. The oscillation frequency is set by the size and temperature of the corona, and thus, the increase in QPO frequency in 1ES\,1927+654 can naturally be explained by a decrease in the size of the corona (see Methods, Section \ref{subsec:corona} for more details). This model also predicts a spectral softening with decreasing coronal size, as is observed in 1ES\,1927+654. However, this model would predict unfeasibly large coronae in RE J1034 and 2XMM J1231, assuming that these QPOs are produced via the same mechanism as in 1ES\,1927+654.

To summarize, the enigmatic AGN 1ES\,1927+654 shows yet another surprise with this persistent, yet rapidly evolving mHz frequency X-ray QPO. Regardless of the QPO origin, such high frequencies imply periodic events from very close to the SMBH, which persist for nearly 2 years. The rapid evolution in the QPO frequency suggests that the QPO could be driven by stable mass transfer from a white dwarf companion, disk instabilities with a dependence on $\dot{M}$, or magnetoacoustic oscillations in a contracting corona. Continued X-ray monitoring is crucial to disentangling these models, as the companion model would suggest a long-lived ($\gtrsim 10$ years) QPO at a similar frequency, while continued correlated changes between the QPO frequency and X-ray flux may instead suggest an instability or oscillation model. An additional test will come in the 2030s, as a $0.1\, M_\odot$ companion would lead to detectable mHz gravitational waves with LISA.

\begin{addendum}
 \item We thank XMM-Newton PI, Norbert Schartel, for approving the target-of-opportunity requests. MM thanks Mason Ng for discussions regarding X-ray timing and Lisa Drummond for discussions about extreme mass ratio models. MM also thanks the organizers and participants of the UCSB KITP TDE Workshop, including, but not limited to, Jane Dai, Giuseppe Lodato, Chris Nixon, Andy Mummery, Alessia Franchini, Itai Linial, and DJ Pasham for their comments, questions, and discussions regarding these results. EK would like to thank Alex Dittmann and Dan Wilkins for discussions. 
 
 RA was supported by NASA through the NASA Hubble Fellowship grant \#HST-HF2-51499.001-A awarded by the Space Telescope Science Institute, which is operated by the Association of Universities for Research in Astronomy, Incorporated, under NASA contract NAS5-26555. 
 AI acknowledges support from the Royal Society. 
 MG is supported by the ``Programa de Atracci\'on de Talento'' of the Comunidad de Madrid, grant number 2022-5A/TIC-24235. 
 CP is supported by PRIN MUR SEAWIND funded by NextGenerationEU. 
 BT acknowledges support from the European Research Council (ERC) under the European Union’s Horizon 2020 research and innovation program (grant agreement number 950533) and from the Israel Science Foundation (grant number 1849/19).
 JW acknowledges support from the NASA FINESST Graduate Fellowship, under grant 80NSSC22K1596.
 This research was supported in part by grant NSF PHY-2309135 to the Kavli Institute for Theoretical Physics (KITP).
 
 \item[Competing Interests] The authors declare that they have no competing financial interests.
 
 \item[Author Contributions] MM and EK led the analysis, interpretation, and manuscript preparation and requested a subset of the XMM-Newton and NICER data. CP and WNA assisted with X-ray timing analysis. JC and KB contributed to the white dwarf accretion modeling and computed the expected LISA signal. JC, RA, MG, and GM provided information about QPEs, their models, and connections to 1ES 1927+654. SL, SBC, ETM, DRS, and OIS triggered two of the XMM-Newton ToOs in this work. ACF suggested the idea of coronal oscillations. AI and JW provided feedback on the timing analysis and relativistic precession model. RAR assisted with comparison to BHXB QPOs and NICER observations. CR, PK, CP, and BT provided feedback on the analysis. CR, IA, ACF, JAG, PK, ML, CP, RAR, and BT assisted the associated XMM-Newton proposal. All authors contributed to the scientific interpretation of the results and provided feedback on the manuscript.
 
 \item[Correspondence] Correspondence and requests for materials should be addressed to MM (email: mmasters@mit.edu).

\item[Data Availability] All the data used in this work are publicly available through XMM-SAS.

\item[Code Availability] The spectra, light curves, and code used to analyze the data and produce all figures have been made publicly available on CodeOcean and GitHub (\url{https://github.com/memasterson/1ES1927_mHzQPO/}).

\end{addendum}

% \putbib[main_bib]
% \end{bibunit}
% \bibliographystyle{naturemag}
% \bibliography{main_bib}

% set thebibliography to continue numbering from the main bib
% \makeatletter
% \apptocmd{\thebibliography}{\setcounter{enumiv}{50}}{}{} 
% \makeatother

\setcounter{figure}{0}
\renewcommand{\figurename}{Extended Data Figure}

% \begin{bibunit}[naturemag]

\begin{methods}

\subsection{Observations \& Data Reduction}
\label{sec:obs}

\subsubsection{XMM-Newton}
\label{subsec:xmm}

XMM-Newton\cite{Jansen2001} has observed 1ES\,1927+654 extensively, including 1 pre-outburst observation in 2011\cite{Gallo2013} and 15 observations since the beginning of its 2018 outburst\cite{Ricci2020,Ricci2021,Masterson2022}. In this work, we focus on the X-ray variability in the 8 newest observations, which were taken between July 2022 and March 2024. The details of these observations are given in Extended Data Table \ref{tab:xmm_obs}; for the first 8 observations, we refer the reader to (\citen{Masterson2022}). We reduced data from the EPIC-PN instrument (Small Window mode, with the thin filter) using the XMM-Newton Science Analysis System (version 20.0.0) with the latest calibration files. We followed standard data reduction procedures, including running \texttt{epproc} to produce calibrated event files, removing periods of significant background flaring, and creating RMF and ARF files with \texttt{rmfgen} and \texttt{arfgen}, respectively. As our main goal was to assess the variability with Fourier techniques, we maximized the amount of continuous data by making a time cut based on when significant background flaring began. No significant background flares were found in the continuous segments. We found only mild evidence for pile-up in a circular extraction region, and we verified that the QPO is still present even if we excise the central core of the PSF. Hence, for timing analysis we used a circular aperture with radius of 35\arcsec, due to the need for high signal-to-noise data to perform X-ray timing. For spectral analysis, which does not require such high count rates, we used an annular aperture with an inner (outer) radius of 10\arcsec\ (40\arcsec) to mitigate any mild pile-up effects. Additionally, for spectral fitting, we binned the spectrum to 25 counts bin$^{-1}$, employed $\chi^2$ minimization for model fitting, and fit the data in the 0.3-10 keV range with a phenomenological model (\texttt{tbabs $\times$ ztbabs $\times$ (zpower + zbbody)} in XSPEC notation), which accounts for both Comptonization (\texttt{zpower}) and a soft excess (\texttt{zbbody}). Finally, we applied corrections to bring the photon arrival times into the solar system barycenter and constructed light curves with 20s time binning. This choice of time binning was motivated by requiring a relatively high signal-to-noise ratio (SNR $\gtrsim$ 4 for 20s binning with a count rate of 1 counts s$^{-1}$), while still being able to resolve rapid variability. A subset of each XMM-Newton light curve is shown in the bottom left panels of Figure \ref{fig:overview} to highlight the dramatic evolution from pre-outburst to now. 

\subsubsection{NICER}
\label{subsec:nicer}

The Neutron star Interior Composition ExploreR (NICER\cite{Gendreau2016}) has been monitoring 1ES\,1927+654 with a cadence ranging from two observations per day to 1-2 observations per week since close to the start of the outburst in May 2018. These high-cadence observations put the outburst and more sparsely-sampled XMM-Newton observations in context. To reduce the NICER data, we ran \texttt{nicerl2} with good time interval (GTI) binning using HEASoft v6.32.1 (NICERDAS version 11). We did not perform any undershoot or overshoot filtering, instead choosing to filter on the background-subtracted products. To estimate the background, we used the \texttt{nibackgen3c50}\cite{Remillard2022}, which provides an empirical estimate of the NICER background based on pre-compiled libraries from empty fields. We filtered on the background-subtracted light curves by employing level 3 filtering\cite{Remillard2022}, with a background-subtracted count rate threshold of 2 cts s$^{-1}$ in the \texttt{s0} band (0.2-0.3 keV) and 0.05 cts s$^{-1}$ in the \texttt{hbg} band (13-15 keV). Additionally, we removed data from detectors 14 and 34, which are known to be noisy, and from any ``hot detectors" using an iterative sigma-clipping procedure that removes data from detectors that have an $E < 0.2$ keV raw count rate that exceed the median $E < 0.2$ keV raw count rate by more than 4$\sigma$. Similarly, we used the same iterative sigma-clipping procedure to remove any GTIs with a 15-18 keV rate that exceeded the median rate for all GTIs by more than 4$\sigma$, as the 15-18 keV events should all be background events. The resulting NICER light curve, which is used primarily to assist in our understanding of the $\sim$day-month timescale variability, is shown in the top left panel of Figure \ref{fig:overview}.

\subsection{Timing Analysis} \label{sec:timing}

To assess the variability, we constructed power spectral densities (PSDs) for each of the new XMM-Newton observations with [rms/mean]$^2$ normalization\cite{Vaughan2003} using the \texttt{pyLag}\footnote{\url{https://github.com/wilkinsdr/pylag}} spectral-timing package\cite{Wilkins2019}. We computed the PSDs from light curves with a time binning of $\Delta t = 20$s. As detailed in Section \ref{subsec:energy_Q_rms}, we utilize the 2-10 keV band primarily for the analysis of the QPO, as this is where the signal is the strongest. Our analysis is done on individual observations (i.e. with continuous light curves) and with the unbinned PSDs.

In February 2023, a telemetry glitch resulted in the loss of 200s of data. As this is a negligible fraction of the total duration of the observation, we linearly interpolate to create a continuous light curve for Fourier analysis. Importantly, the QPO is still present if we only analyze the data before the glitch; that is, this linear interpolation has no significant impact on the QPO feature. Due to limited visibility of the source in July-August 2022, four short ($\lesssim$ 25 ks) observations were taken within one week of one another (ObsIDs 0902590201, 0902590301, 0902590401, 0902590501). Motivated by their close temporal proximity and in an effort to increase the signal-to-noise, we analyzed these observations simultaneously (i.e. through joint fitting with parameters tied; see details in Section \ref{subsec:significance}). Where we show binned PSDs (for visual purposes only), we computed an unbinned PSD for each observation, binned these across the four observations by averaging at each frequency, and then averaged over neighboring frequencies. In March 2024, we obtained two observations (ObsIDs 0932392001, 0932392101), again taken within roughly one week. The first shows a clear detection of the QPO, while the second only suggests a marginal detection (at a very similar frequency as the first observation). We analyze these observations separately, but show binned PSDs together in Figure \ref{fig:lc_psd} and Extended Data Figure \ref{fig:PSD_energy} (with the same methodology as for the July-August 2022 data). 

\subsubsection{Broadband Noise Models} \label{subsec:broadband}

Estimating the significance of the QPO requires an accurate understanding of the underlying broadband noise, as past work has shown that QPO significance can be overestimated if this is improperly modeled\cite{Vaughan2005b,Gonzalez-Martin2012}. The broadband noise of most AGN can often be described by a power-law with $P(f) \propto f^{-\alpha}$. The slope tends to be characteristic of red noise with $\alpha = 2$, but on longer timescales, there is often a break to a shallower slope with $\alpha \approx 1$\cite{Uttley2002,Markowitz2003,McHardy2004}. The frequency of this break between the two regimes has been shown to correlate strongly with the black hole mass and also weakly with the accretion rate\cite{McHardy2006}. However, in the case of 1ES\,1927+654, the high energy (2-10 keV) PSDs show a much shallower slope close to white noise ($\alpha = 0$). Thus, we explored only two broadband noise models for the 2-10 keV PSDs -- a power-law with a free index, $\alpha$, and a Lorentzian with the centroid frequency fixed at $f = 0$ (commonly used in modeling the broadband noise of BHXBs, e.g. ref. \citen{Belloni2002}). For both of these models, we included a constant term to account for the Poisson noise, which dominates at high frequencies. 

To assess the broadband noise, we fit the unbinned PSDs using the log likelihood given by 
\begin{equation} \label{eqn:likelihood}
    \log\mathcal{L} = -\sum_{j} \left(\frac{I_j}{S_j} + \log S_j\right),
\end{equation}
where $I_j$ is the observed power, $S_j$ is the model power, and $j$ corresponds to the $j$th frequency bin. This log likelihood utilizes the fact that the unbinned periodogram follows a $\chi^2_2$ distribution (i.e. a $\chi^2$ distribution with 2 degrees of freedom; see Equation~(17) in ref. \citen{Vaughan2010}). Fitting the unbinned PSDs is necessitated by the low number of frequency points in AGN PSDs; specifically, we are unable to bin the PSDs to the Gaussian regime ($N \gtrsim 25$ points per bin) without smoothing out the QPO significantly. To fit the PSDs, we implement a Bayesian approach, first performing a maximum likelihood fit to the data, followed by a Markov Chain Monte Carlo (MCMC) analysis using the results of the maximum likelihood fit to initialize walkers. We utilize the \texttt{emcee} package\cite{Foreman-Mackey2013} to fit the data, using 32 walkers with 55,000 chain steps per walker followed by a burn-in of 5,000 steps per walker. Each resulting chain was inspected visually to check for convergence. 

Extended Data Figure \ref{fig:broadband} shows the PSDs from the observations we analyzed individually, fit with both the power-law and Lorentzian broadband noise models. For the four observations from July-August 2022, we fit with the same methods outlined above, but fit all four observations simultaneously with all model parameters tied due to the weakness of the QPO (see Section \ref{subsec:significance} for further details). The unbinned PSDs and resulting model fit for this simultaneous analysis are shown in Extended Data Figure \ref{fig:aug2022_broadband}. Both models provide a sufficient fit to the data, although the Lorentzian model does a slightly better job of fitting the lowest frequencies. It is evident from these fits that there is an additional peak in the PSD that is not captured in either broadband noise model. This QPO feature shows up near 0.9, 1.7, 2.2, and 2.3 mHz, respectively across the July-August 2022, February 2023, August 2023, and March 2024 data. In each observation, the QPO is clearly resolved in the unbinned PSD, showing up as an excess of power in multiple frequency channels, thereby adding to our confidence that the feature is real. In March 2024, the first observation (ObsID 0932392001) shows a significant detection, while the second observation (ObsID 0932392101) taken 8 days only shows a weak excess around the same frequency. Further monitoring will be necessary to tell whether there is short term variability in the QPO amplitude or if the QPO is turning off.

\subsubsection{Estimating QPO Significance} \label{subsec:significance}

We estimated the significance of the QPO in our observations through two independent methods, one focused on the fits with and without an additional Lorentzian component for the QPO and another with a sample of simulated light curves. First, we performed a fit to the unbinned data with an additional Lorentzian component to account for the QPO feature. We then used the Akaike Information Criterion (AIC)\cite{Akaike1974} to test whether this model provided a significantly better fit than the broadband noise model alone. Crucially, AIC does not require nested models, as is required in the commonly-used likelihood ratio test. We use the sample-corrected AIC\cite{Sugiura1978,Emmanoulopoulos2016}, which is defined as 
\begin{equation}
    \mathrm{AIC_c} = 2k - 2C_\mathrm{L} - 2\log\mathcal{L} + \frac{2k(k+1)}{N-k-1},
\end{equation}
where $C_\mathrm{L}$ is the (unknown) likelihood of the true model, $k$ is the number of free parameters in the model, $N$ is the number of data points, and $\mathcal{L}$ is the maximum likelihood, as defined by Equation~(\ref{eqn:likelihood}). With the difference between the AIC values for two models ($\Delta \mathrm{AIC} = \mathrm{AIC_{c,1}} - \mathrm{AIC_{c,2}}$), we can then estimate the relative likelihood, which is given by
\begin{equation}
    p_\mathrm{AIC} = e^{-\Delta \mathrm{AIC}/2},
\end{equation}
and, importantly, is independent of the unknown $C_\mathrm{L}$. If $\Delta \mathrm{AIC}$ is large ($\gtrsim 10$), then $p_\mathrm{AIC}$ is small ($\lesssim 0.01$) and the model with the smaller AIC is strongly favored. To estimate the QPO significance, we assumed Model 2 is the broadband plus QPO model and Model 1 is the broadband only model. The resulting $\Delta \mathrm{AIC}$ are quite large (most $> 30$), with corresponding $p_\mathrm{AIC}$ values of $< 10^{-4}$, irrespective of the choice of broadband noise model. Hence, the QPO is highly statistically significant. The respective $p_\mathrm{AIC}$ values for each observation and broadband noise model are given in Extended Data Table \ref{tab:qpo}.

As another estimate of the significance of the QPO, we estimated how often the observed peak would occur in a randomly generated light curve with the same broadband noise parameters. To implement this, we simulated observations based on the broadband noise models using the method outlined by (\citen{Timmer1995}). For each observation and each broadband noise model, we generated $10^5$ simulated light curves with the same sampling, mean, and RMS as the observed light curve. For each simulation, we drew parameters for the broadband model randomly from the posterior distributions from our MCMC analysis, thereby taking into account the uncertainty in the broadband noise modeling. Then, for each simulated light curve, we computed the PSD and proceeded with the same analysis that we performed on the observed data (i.e. fitting with maximum likelihood and then running an MCMC). 

To assess the significance of the QPO, we compared statistics of the simulated light curves to those of the observed data. Most QPOs in AGN are unresolved in the unbinned PSD, showing up in only one frequency channel\cite{Gierlinski2008,Alston2014}. Thus, many statistics for assessing the QPO significance that depend only on a single frequency channel will underestimate the significance of the QPO in 1ES\,1927+654, which appears in multiple consecutive frequency channels. For example, a commonly used statistic is $T_R = \max_j R_j$, where $R_j = 2 I_j / S_j$ and $I_j$, $S_j$ are the observed and model powers, respectively. This statistic depends only on a single peak in the unbinned PSD, thereby missing the added significance that comes from having more than one frequency channel with excess power around the same frequency. Instead, we utilize the traditional $\chi^2$ statistic or summed square error (SSE) estimator\cite{Vaughan2010}, which is given by
\begin{equation}
    T_\mathrm{SSE} = \sum_{j} \left(\frac{I_j - S_j}{S_j}\right)^2.
\end{equation}
This statistic is well-suited for the case of the QPO in 1ES\,1927+654, as all of the data points contribute to $T_\mathrm{SSE}$, rather than just a single point. However, it does not take into account the fact that the QPO appears as an excess in consecutive frequency channels. Thus, we except the SSE estimator to provide a lower bound on the true QPO significance; hence, where we quote a single significance value, we report the AIC significance (for the lower of the two broadband noise models).

We perform each of these tests on individual observations, with the exception of the data from July-August 2022. Due to limited visibility in July-August 2022, a total of four observations were taken over one week for a total of 66 ks, averaging just over 16 ks per observation (after data cleaning). These observations show a hint of a QPO at $\sim 0.9$ mHz, but the QPO does not appear significantly in any individual observation, likely due to the limited number of cycles per observation. To assess the potential QPO feature, we fit all 4 observations simultaneously with the same broadband noise models described in Section \ref{subsec:broadband}. We computed the log likelihood for each observation and then added these together to compute the total log likelihood (i.e. we did not bin the data before fitting). This fit then provided the baseline for computing various statistics for estimating significance. Extended Data Figure \ref{fig:aug2022_broadband} shows the resulting broadband fit, with a consistent excess seen around 0.9 mHz in all four observations. 

To estimate the significance of this feature, we use the same estimators described above, but this time use all four observations simultaneously to improve our limited statistics. The SSE estimator does not show a significant detection of the QPO ($p \sim 0.5$); however, since the SSE estimator is a measure of the global goodness-of-fit, it does not account for the fact that the excess is seen at roughly the same frequency in each of these four observations. Instead, we expect that the AIC estimator is more accurate for this set of observations, as it can properly account for this QPO-like feature showing up at the same frequency. To measure the AIC significance, we again fit these four observations simultaneously, with all of the parameters fixed across all observations, including the frequency, width, and normalization of the Lorentzian used to fit the QPO. The difference in AIC between the broadband only model versus broadband with QPO model suggests that the QPO is significant at around 4$\sigma$ significance. Thus, when combined with the clear detection of the QPO at later times, we suspect that there is a broad QPO in these 2022 observations. 

In Extended Data Table \ref{tab:qpo}, we report the significance of the QPO for each of these methods, which we estimate assuming both of the broadband noise models. The QPO is highly significant ($p \lesssim 0.01$) for almost all of the recent observations, irrespective of which broadband noise model or significance test we use. Only the second of the two most recent observations from March 2024 (ObsID 0932392101) and the group of four observations from July-August 2022 (ObsIDs 0902590201-501) show marginally detected features ($\lesssim 3\sigma$). On average the AIC test gives higher significance, but both methods point to a highly significant QPO in 1ES\,1927+654. We emphasize that the analysis is done on each epoch separately and is still highly significant; combining these findings across all four epochs between 2022 and 2024 would only lead to even higher significance.

\subsubsection{QPO Properties: Energy Dependence, Coherence, and Fractional RMS} \label{subsec:energy_Q_rms}

To assess the energy dependence of the QPO in 1ES\,1927+654, we computed PSDs across a number of different energy ranges. For all of the observations, the QPO is present and at roughly the same frequency in all energy ranges, but, as in BHXBs, it appears to be the strongest in the higher energy bands (see Extended Data Figure \ref{fig:PSD_energy}). Additionally, the lower energies have significantly higher broadband noise at low frequencies, which can be attributed to the typical red noise variability seen in AGN. The 2-10 keV band does not show as strong of a low frequency noise component and therefore makes analysis of the QPO significance more robust in this band. Additionally, the fractional RMS of the QPO increases with increasing energy, most notably from 2-7\% in 0.3-2 keV, up to 15-19\% in 2-10 keV (see Extended Data Figure \ref{fig:rms}). We computed these values by fitting the QPO with an additional Lorentzian component on top of the broadband noise, and we find that all of the observations with a significant QPO detection give the same trend. These values are given in Extended Data Table \ref{tab:qpo} and are comparable to the QPOs around SMBHs\cite{Alston2014,Pasham2019,Ashton2021} and both HFQPOs and Type C LFQPOs in BHXBs\cite{Remillard2006a,Ingram2019}.

The QPO is clearly resolved in the unbinned PSDs, and hence, we estimated the coherence by adding an additional Lorentzian component to our power-law model for the broadband noise and fitting for the centroid frequency and width. From this fit, we compute the quality factor, $Q = f_\mathrm{QPO}$/FWHM$_\mathrm{QPO}$, finding $Q = 8.7, 8.0, 10.1$ in the February 2023, August 2023, and March 2024 observations respectively. Again this is comparable to what is seen in QPOs around SMBHs\cite{Alston2014,Pasham2019} and BHXBs\cite{Remillard2006a,Ingram2019}, and suggests a highly coherent signal, as can be seen from the light curves even by eye. The clear exception to these high $Q$ values is the 2022 observations, which show a relatively broad feature with $Q \approx 2.3$. However, it is not clear whether the breadth of this feature is intrinsic or due to some short timescale variability in the QPO frequency that is being smeared out by considering four observations over the span of a week. However, the two observations from March 2024, separated by 8 days, suggest that there may be some variability in the QPO on short timescales. We will investigate the short-term evolution of the QPO with future XMM-Newton observations.

\subsection{SMBH Mass and Spin Constraints from the QPO} \label{subsec:mass_spin}

If the physical origin of the QPO is well-understood, then the frequency can provide constraints on both the mass and spin of the SMBH. In this section, we provide the mass and spin limits that arise from associating the QPO with the orbital timescale at the ISCO, which would hold in both the case of a circularized orbiter or disk precession models for the QPO. It is, however, important to note that these limits may not hold in other QPO models (e.g. if the QPO is produced in the corona or jet and not constrained by the orbital timescale at the ISCO).

The mass estimate is of particular importance for 1ES\,1927+654, as the mass has been previously debated and is crucial to interpreting the super-Eddington nature of the outburst. Two initial estimates, based on the narrow [O III] $\lambda$5007 line\cite{Tran2011} and the width of the newly-formed broad H$\beta$ line at the beginning of the outburst\cite{Trakhtenbrot2019}, suggested $M \approx$ few $\times \, 10^7 \, M_\odot$. However, these methods have large scatter and assume that the lines are virialized, which is especially problematic for the analysis of the broad H$\beta$ line as it had only just formed at the beginning of the 2018 outburst. Instead, recent analysis of the host galaxy suggests a mass of $M \approx 10^6 \, M_\odot$, from $M_\mathrm{BH}-\sigma_*$, $M_\mathrm{BH}-M_\mathrm{bulge}$, $M_\mathrm{BH}-M_*$ relations\cite{Li2022}.

The fastest frequency that can naturally arise from a test particle in orbit around the SMBH at a given radius is the orbital frequency, given by
\begin{equation}
    f_\phi = \pm \frac{c^3}{2\pi GM} \left(\frac{1}{r^{3/2} \pm a}\right),
\end{equation}
where $a$ is the dimensionless spin parameter of the SMBH, $r$ is the radius in terms of gravitational radii ($r_g = GM/c^2$), and $M$ is the mass of the SMBH. Using the highest frequency QPO, $f_\mathrm{QPO} = 2.34$ mHz in March 2024, with a maximally spinning SMBH ($a = 0.998$) yields an upper limit on the mass of the SMBH of $M \leq 5.8 \times 10^6 \, M_\odot$. This constraint could be increased to twice this value ($M \leq 1.16 \times 10^7 \, M_\odot$) if the QPO corresponds to twice the orbital frequency, which could be the case if the QPO is produced by a companion striking the disk twice per orbit. It is possible that the QPO is coming from further from the SMBH than the ISCO, assuming that the mass is significantly less than $M \approx 10^6 \, M_\odot$. However, given that the source reaches a bolometric luminosity of $L_\mathrm{bol} \approx 5 \times 10^{44}$ erg s$^{-1}$ during the outburst\cite{Li2024}, such a scenario would require an extremely high accretion rate ($L \gtrsim 10 L_\mathrm{Edd}$) for an SMBH with $M \lesssim 5 \times 10^5 \, M_\odot$. Together, this evidence points to a SMBH of $M \approx 10^6 \, M_\odot$, consistent with the host galaxy scaling relations\cite{Li2022}.

Assuming a mass of $1.38 \times 10^6 \, M_\odot$, the best SMBH mass estimate\cite{Li2022}, we can also place a lower limit on the spin of the SMBH by assuming that the QPO must be produced outside of the ISCO. Extended Data Figure \ref{fig:isco} shows the location of the QPO in terms of the ISCO for varying spins. The most rapid QPO, with a frequency of 2.34 mHz, gives $a > 0.43$, implying that 1ES\,1927+654 is a spinning SMBH. This result depends sensitively on the mass estimate, which has significant uncertainty and comes primarily from host galaxy estimates. The shaded regions of Extended Data Figure \ref{fig:isco} show how the uncertainty on the mass of the SMBH influences our spin constraint; it is possible to get a non-spinning SMBH if the mass is on near the lower limit from host galaxy scaling relations. However, as there is no significant broad iron line in this source, the QPO may be one of the only ways to place any limits on the spin of this enigmatic SMBH. 

\subsection{Comparison to QPOs in Black Hole X-ray Binaries} \label{subsec:bhxbs}

Black hole X-ray binaries (BHXBs), consisting of a stellar mass black hole accreting from an orbiting companion star, exhibit strong QPOs, ranging in frequency from mHz up to several hundreds of Hz\cite{Remillard2006a,Ingram2019}. These QPOs are far more abundant than those around SMBHs, likely due to the relatively long timescales required to probe QPOs around SMBHs. In the BHXB literature, the low-frequency QPOs (LFQPOs; $f \lesssim 30$ Hz) are broken into three classes -- Type A, B, and C -- based on their strengths, widths, and when they occur during the outburst. Type C QPOs, typically found in the low-hard and hard-intermediate states, are the most common type of LFQPO, and they are characterized by relatively narrow peaks ($Q \gtrsim 8$) and high fractional RMS ($\lesssim 20$\%). The frequency of Type C QPOs are tightly correlated with the photon index of the power-law component, which has led to the suggestion that they arise from precession of a hot inner flow\cite{Ingram2009}. This geometric origin of Type C QPOs is also supported by inclination dependence on QPO properties\cite{Heil2015,Motta2015,vandenEijnden2017} and modulation of the iron line strength with QPO phase\cite{Ingram2016,Nathan2022}. Type B QPOs, on the other hand, mark the transition to the soft-intermediate state and are likely associated with the launching of a transient jet ejection\cite{Soleri2008,Homan2020}. They show a relatively narrow range of QPO frequencies ($\approx 1-6$ Hz), occur when the broadband spectrum shows only weak red noise, and show a strong correlation with the power-law flux\cite{Motta2011}. Interestingly, the QPO in 1ES\,1927+654 is seen in a state with very little red noise in the power-law component, which is akin to Type B QPOs, but rather unusual for standard AGN\cite{Uttley2002} (although note that QPEs also show very low levels of variability in their quiescence\cite{Miniutti2019,Giustini2020}). Similarly, 1ES\,1927+654 also showed a recent rise in radio flux, potentially from newly launched jet\cite{Meyer2024} (consistent with launching in mid-2022; S. Laha et al. ApJ, submitted) and akin to what is seen in the state transitions in BHXBs.

While LFQPOs are nearly ubiquitous among BHXBs in outburst, only a small subset also show high-frequency QPOs (HFQPOs; $f \gtrsim 60$ Hz)\cite{Morgan1997,Remillard1999a,Remillard1999b,Homan2001,Miller2001,Strohmayer2001,Remillard2002,Belloni2012}. HFQPOs tend to be seen almost exclusively in the high-soft state, are fairly constant in frequency even across different outbursts, and have RMS $\approx$ 1-10\% and $Q \approx 5-30$. Importantly, both LFQPOs and HFQPOs primarily occur in the Comptonized component, with fractional RMS increasing significantly as a function of energy. A powerful constraint on models for HFQPOs are the existence of harmonics, which are commonly observed at a 3:2 frequency ratio\cite{Strohmayer2001,Remillard2002,Motta2014a}. A number of models have tried to explain this observational property, with, for example, resonances between the orbital, vertical epicyclic, and radial epicyclic frequencies\cite{Abramowicz2001,Abramowicz2003} and trapped pressure oscillations in a thick inner flow\cite{Rezzolla2003}. The PSDs for 1ES\,1927+654 show some features that hint at potential harmonics, but none of these features are statistically significant, nor do they exist at the 3:2 resonance observed in HFQPOs.  

Taking the SMBH in 1ES\,1927+654 to be roughly $10^6 \, M_\odot$, a 2 mHz QPO would scale up to around 200 Hz in a $10 \, M_\odot$ BHXB, making this QPO similar to the HFQPOs in frequency. The QPO in 1ES\,1927+654 is narrow ($Q \approx 10$), has a high RMS (RMS $\approx 15$\%) in 2-10 keV, and occurs primarily in the Comptonized component, and thus the properties are roughly consistent with the properties of HFQPOs in BHXBs. However, the most notable difference between HFQPOs and the QPO in 1ES\,1927+654 is the frequency evolution, as HFQPOs are remarkably stable in frequency, even across order of magnitude changes to the X-ray flux\cite{Remillard2002,Belloni2012}. This suggests that the HFQPOs are set by the fundamental properties of the black hole (e.g. mass and spin, which set the surrounding spacetime), rather than more complex and variable accretion physics. In this way, the QPO in 1ES\,1927+654 is quite different from HFQPOs; it has a strong dependence on the X-ray flux and spectral shape (see Extended Data Figure \ref{fig:qpofreq_fluxevol}). While this behavior is similar to LFQPOs, it is quite difficult to get the typical LFQPO models (e.g. precession\cite{Ingram2009}) to work for 1ES\,1927+654 due to how close the QPO arises relative to the ISCO. 

Therefore, a direct comparison between 1ES\,1927+654 and a specific type of QPO in BHXBs is quite difficult; clearly, the collective phenomena do not match any designation perfectly. This makes 1ES\,1927+654 unique among the SMBHs showing strong QPOs, as these tend to be quite stable and are thought to be HFQPO analogs\cite{Gierlinski2008,Alston2014,Alston2015,Pasham2019}. Despite such a high frequency, the flux dependence of the QPO frequency suggests a connection to the accretion flow and X-ray corona, as seen in Type B and C QPOs. We explore the connection to the physics thought to govern these LFQPOs in Sections \ref{subsec:instability} and \ref{subsec:corona}.

\subsection{Comparison to Quasi-Periodic Eruptions (QPEs)} \label{subsec:qpes}

Quasi-periodic eruptions (QPEs\cite{Miniutti2019,Giustini2020,Arcodia2021,Chakraborty2021,Arcodia2024a}) are a new class of nuclear transients that show repetitive soft X-ray flares. These flares reach up to $L_{X,\,\mathrm{flare}} \approx 10^{43}$ erg s$^{-1}$ and recur on timescales of hours to days, which corresponds to the orbital timescales at roughly 40-200 $R_g$. All of the known QPEs show super-soft, thermal X-ray spectra in quiescence ($kT_\mathrm{quiescent} \approx 50-80$ eV, where detected) and become hotter during their flares ($kT_\mathrm{flare} \approx 100-200$ eV). 

To date, there have been two main classes of models used to explain QPEs: (1) accretion disk instabilities\cite{Raj2021,Pan2022,Kaur2023,Pan2023,Sniegowska2023}, and (2) a stellar-mass ($\lesssim 100\, M_\odot$) companion. This second class of models can be broken down into two subclasses of models, namely those in which the companion interacts with an underlying accretion disk\cite{Dai2013,Sukova2021,Xian2021,Franchini2023,Linial2023,Tagawa2023} and those in which the companion mass transfers onto the SMBH\cite{King2020,King2022,Krolik2022,Metzger2022,Zhao2022,King2023}. Timing variations, most prominently a common short-long recurrence pattern across almost all sources\cite{Arcodia2022,Miniutti2023b}, favor the impacting orbiter models, as this can be naturally explained by a slightly elliptical orbit where the collision timescale is longer on one side of the disk than the other. Similarly, the recent finding that Lense-Thirring precession of the underlying disk could explain the more chaotic recurrence timing variations in eRO-QPE1 gives additional support to the impacting orbiter models\cite{Chakraborty2024}. Assuming QPEs are caused by an orbiting companion punching through the disk, the typical timescales imply a companion at roughly 100 $R_g$ from the SMBH. In this regime, the gravitational wave timescale is long ($\approx 10^5$ years), implying that there should be little observed secular evolution of the period from gravitational wave emission.

1ES\,1927+654 shows many similarities to QPEs, both in terms of the flaring properties (e.g. soft X-ray spectra, quasi-periodicity) and the properties of the host galaxies and SMBHs (e.g. low SMBH mass, connections to TDEs, and super-soft AGN intrinsically lacking a broad line region). The timescale, amplitude, and duty cycle of the QPO in 1ES\,1927+654, however, differ from QPEs. The QPO period of 7-18 minutes in 1ES\,1927+654 corresponds to orbital timescales at $\lesssim 10 \, R_g$, which is well within the tidal radius of a solar-like companion that has been invoked to explain QPEs\cite{Linial2023}. The QPO has an amplitude on the order of 15\% in 2-10 keV (see Extended Data Figure \ref{fig:rms}), which is significantly lower than the order of magnitude variability seen in QPEs, but it is important to note that this is energy-dependent and the underlying accretion rate is likely different in these two systems. The baseline X-ray luminosity for 1ES\,1927+654 is at least an order of magnitude higher than most QPEs; GSN 069 is a notable exception, as it is significantly brighter than most other QPEs, but it only shows QPEs below $L \approx 0.4 \, L_\mathrm{Edd}$\cite{Miniutti2023a}. Additionally, 1ES\,1927+654 does have an X-ray corona that, although soft compared to standard AGN, produces emission out to $\gtrsim$ 10 keV, unlike in QPEs. Finally, the QPO in 1ES\,1927+654 does not show any obvious signs of ellipticity, in contrast to the short-long behavior seen in QPEs, which could arise from the circularization of the orbit as the companion approaches the SMBH. Similar comparisons have been drawn between QPEs and a QPO in the TDE candidate 2XMM J1231, leading various authors to question whether QPOs and QPEs are truly distinct or potentially different observational manifestations of similar physics\cite{King2023,Webbe2023}. For example, it has been suggested that both QPOs and QPEs can result from a companion mass transferring at pericenter, where oscillations are seen if the viscous timescale is longer than the orbital period, and vice versa\cite{King2023,Miniutti2023b}. Thus, while there are differences, the suggestive similarities between QPEs and the QPO in 1ES\,1927+654 warrant investigating whether they can be driven by similar physical models (see Section \ref{sec:models}).

It is also interesting to note that two of the bonafide QPEs show potential QPOs as well. In the QPE source RX J1301.9+2747, the potential QPO has a period that is significantly shorter than the QPE recurrence time ($P_\mathrm{QPO} \approx 0.4$ hours, $t_\mathrm{QPE,\, recur} \approx 3-5$ hours), and its period appears to be stable over more than 18 years\cite{Song2020}. This stability, and the similarity to both RE J1034 and HFQPOs in BHXBs, lead to the suggestion that this QPO may be related to disk resonance modes, rather than a direct connection to the QPEs\cite{Song2020}. On the other hand, the QPE discovery source GSN 069 shows a potential QPO at the same period as the QPE recurrence time, even across multiple epochs\cite{Miniutti2023a,Miniutti2023b}. This suggests that the QPO is directly linked to the driving mechanism of the QPEs (e.g. fluctuations in the mass accretion rate as a result of shocks producing the QPEs). It is important to note that the lack of prominent QPE-like bursts in 1ES\,1927+654 does not preclude the QPO from being driven by similar physics. If, for example, the companion is a compact object, then the luminosity of QPEs depends on the relative velocity between the orbiter and the disk at the point of impact\cite{Franchini2023}, and thus, a companion on either a retrograde or a more inclined orbit would have lower luminosity QPEs that may be missed below the underlying accretion luminosity. Ultimately, the existence of both QPEs and QPOs in low mass, super-soft X-ray AGN motivates further exploration of their potential connection.

\subsection{Potential Models for the QPO} \label{sec:models}

\subsubsection{Extreme Mass Ratio Models} \label{subsec:EMR}

The recent discovery of QPEs has garnered increased interest in the electromagnetic counterparts to Extreme Mass Ratio Inspirals (EMRIs), as they help constrain rates\cite{Arcodia2024b} and signify a clear channel to multi-messenger signals in the mHz regime. The striking similarities between QPEs and 1ES\,1927+654 motivate considering a similar model for the QPO. The most notable difference -- the timescale of periodicity -- actually means that the companion would be much closer to the SMBH and would therefore be more likely to produce a detectable gravitational wave signal than QPEs. In this section, we discuss how an extreme mass ratio companion ($M_\mathrm{companion} \lesssim 10\, M_\odot$, $q = M_\mathrm{companion} / M_\mathrm{BH} \lesssim 10^{-5}$) could produce the observed QPO.

The frequency evolution of the QPO, which is decelerating with time (i.e. negative $\ddot{f}$), can constrain the nature of a potential companion. This evolution clearly deviates from the expected evolution of an inspiraling companion evolving solely due to the emission of gravitational waves (GWs), which should accelerate with time. To highlight this, we computed the expected evolution of the orbital frequency from the emission of GWs, using the leading-order frequency evolution for two point particles\cite{Peters1964}, given by
\begin{equation}
    \dot{f} = \frac{96 G^{5/3}}{5 c^5} \frac{M_1 M_2}{\left(M_1 + M_2\right)^{1/3}} \left( 2\pi \right)^{8/3} f^{11/3},
\end{equation}
where $f$ is the orbital frequency, $M_1$ is the mass of the SMBH, and $M_2$ is the mass of the orbiter. Here we assumed that the companion is on a circular orbit and that the QPO frequency corresponds to the orbital frequency of the companion, motivated by efficient eccentricity and inclination damping predicted for ``wet" EMRIs (i.e. those embedded in gaseous disks)\cite{Pan2021}. As our initial condition, we take the most highly significant QPO with the best constraint on the frequency, which is from February 2023 with frequency of $1.67 \pm 0.04$ mHz. Figure \ref{fig:qpo_freq_evol} shows the comparison between the observed QPO frequency evolution (black) and evolution due to GR alone for various companion masses (color). GR alone cannot reproduce the observed frequency evolution, even if we assume that the orbit has a relatively high eccentricity (as this only speeds up the frequency evolution, but still leads to an accelerating frequency increase). Additionally, the GR models would predict that the QPO should be in an observable band ($\approx$ mHz) in previous XMM-Newton observations, and we do not see such a feature in observations up to January 2021. Thus, if we are going to invoke an extreme mass ratio companion to explain the QPO in 1ES\,1927+654, there needs to be additional physics at play.

There are two main additional forces that could be invoked to explain the frequency evolution of the QPO in 1ES\,1927+654 -- gas drag, which would likely also speed up the rate of inspiral due to dissipation of angular momentum during interactions between the companion and the disk, and stable mass transfer from the companion onto the SMBH, which could in principle slow down the inspiral rate due to the outward transfer of angular momentum from the accretion disk or stream. Assuming that the companion fills its Roche lobe at the minimum observed period of the QPO of 7.1 minutes, the required density implies object must be a white dwarf (WD) with a mass of $M_\mathrm{WD} \gtrsim 0.1 \, M_\odot$\cite{Eggleton1983}, where the lower limit arises from assuming a zero-temperature WD mass-radius relation\cite{Verbunt1988}. It is important to note that in order for angular momentum to be transferred back to the companion efficiently, the accreted material should form a disk before plunging into the SMBH. 

To explore the feasibility of this model, we can investigate how these angular momentum sinks and sources affect the orbit. Following white dwarf accretion literature\cite{Marsh2004}, and for the time being neglecting the effects of gas drag, we can write the rate of change of the frequency as
\begin{equation}
    \frac{\dot{f}}{f} = -3\frac{\dot{J}_\mathrm{orb}}{J_\mathrm{orb}} + 3 \frac{\dot{M}_\mathrm{WD}}{M_\mathrm{WD}}\left(1 - q\right),
\end{equation}
where the change in orbital angular momentum can be decomposed into the effects of GR and mass transfer. Ultimately, this gives
\begin{align}
    \frac{\dot{f}}{f} &= -3\frac{\dot{J}_\mathrm{GR}}{J_\mathrm{orb}} + \frac{\dot{M}_\mathrm{WD}}{M_\mathrm{WD}}\left[3\left(1 - q\right) + \sqrt{1 + q}\right] \\
    &= \frac{96 G^{5/3}}{5 c^5} \frac{M_1 M_2}{\left(M_1 + M_2\right)^{1/3}} \left( 2\pi f \right)^{8/3} + \frac{\dot{M}_\mathrm{WD}}{M_\mathrm{WD}}\left[3\left(1 - q\right) + \sqrt{1 + q}\right],
\end{align}
for which we refer the reader to (\citen{Marsh2004}) for more details. Without any additional angular momentum sinks, a turnaround ($\dot{f} = 0$) at $f = 2.34$ mHz with $M_\mathrm{WD} = 0.1 \, M_\odot$ and $M_\mathrm{BH} = 10^6 \, M_\odot$ requires the WD to be losing mass at a rate of $\approx 3 \times 10^{-4} \, M_\odot$ year$^{-1}$, which is higher than observed accreting WDs\cite{Ramsay2018,Strohmayer2021}, but near the theoretical expectation for WDs at the boundary between inspiraling and outspiraling\cite{Marsh2004}. This phase is short lived, potentially explaining the lack of observed systems at this high accretion rate. 

From Figure \ref{fig:qpo_freq_evol}, we can see that to drive the rapid evolution in frequency between February 2023 and August 2023 with a circular orbit, we would require a companion with $M \approx 7.5 M_\odot$, which is much larger than the $0.1 \, M_\odot$ WD. Thus, for the stable mass transfer model to work, we require either an eccentric orbit (which increases $\dot{f}_\mathrm{GR}$) or another angular momentum sink in the system, such as gas drag. Adding in gas drag is rather uncertain, as this depends strongly on both the disk surface density and the cross-section of interaction for the companion. For stars, this is often simply their size, whereas for compact objects, this is often taken to be the Bondi-Hoyle radius\cite{Linial2024}. To provide an order of magnitude estimate, we compared the observed $\dot{f}$ between February-August 2023 with the expected $\dot{f}$ from GWs  for a 0.1 $M_\odot$ WD in a circular orbit, finding that gas drag (or any additional angular momentum sink) must provide roughly a factor of 40 times higher $\dot{f}$ to match the observed evolution. When combining the effects of gas drag with the mass transfer to offset the angular momentum losses, the higher $\dot{f}$ requires a higher mass transfer rate from the WD of $\dot{M} \approx 10^{-2} \, M_\odot$ year$^{-1}$ to stall the orbiter at the observed $f = 2.34$ mHz in March 2024. This brings into question whether the orbiter could survive for an extended period of time, but the orbiter is likely embedded in the disk and therefore continuously fed with material. 

This simplified picture has only included the effects of mass transfer, gas drag, and GW emission, neglecting potential additional sources of angular momentum. The high Eddington rate in 1ES\,1927+654 means that the accretion disk is likely not a simple thin disk, and early X-ray observations showed evidence for a relativistic outflow launched in the geometrically thick inner accretion flow\cite{Masterson2022}. The complex interplay between the EMRI and outflows, turbulence, and magnetic fields in the disk can also dissipate energy and angular momentum. Additionally, theoretical work has shown that migration traps (produced by differential rotation) can exist near the ISCO in slim accretion disks\cite{Peng2021}, albeit for more massive SMBHs than 1ES\,1927+654. Thus, there is additional, complex physics at play that can also drive the EMRI to evolve in the observed manner, but a detailed analysis is beyond the scope of this work. 

As it was suggested that the initial outburst and destruction of the X-ray corona could be caused by a TDE\cite{Trakhtenbrot2019,Ricci2020,Ricci2021}, it is tempting to connect this companion model for the QPO to the initial 2018 outburst through a partial TDE (pTDE) leaving behind a core that becomes the companion. pTDEs are commonly invoked to explain repetitive nuclear flares on timescales of 100s of days\cite{Payne2021,Wevers2023,Somalwar2023}, but there are limits to how bound a single star can become to the SMBH due to kicks from the interplay between tides and asymmetric mass loss\cite{Gafton2015,Cufari2023}. For shorter period transients, like QPEs, this has lead to the suggestion that either the orbiter pre-dates the TDE\cite{Linial2023} or that other dynamics are involved (e.g. Hills capture of a binary)\cite{Cufari2022}. With the Hills mechanism, it would be possible to get a star on an orbit of $\lesssim$ days, if the initial binary was tightly bound (e.g. separations of $a_\mathrm{bin} \lesssim 10^{-3}$ AU). Additional constraints on the orbital dynamics of the TDE in 1ES\,1927+654 may be able to be inferred from systematic shifts of the broad Balmer lines during the outburst, which are consistent with an eccentric ($e \approx 0.6$) orbit, albeit with much longer periods than we see for the QPO ($\tau_\mathrm{dyn} \approx 1100$ days)\cite{Li2022}. In 1ES\,1927+654, the presence of an existing accretion disk complicates the simple dynamics usually studied in pure TDEs, as the disk will dissipate additional energy and angular momentum to bring a remnant closer to the SMBH. However, the four year timescale from the initial outburst to the production of QPOs at $\lesssim 10\, R_g$ is still rather fast. It has also recently been suggested that disk-captured TDEs in AGN could produce the short-period orbits seen in QPEs\cite{Ryu2024,Wang2024}. Thus, while it may be possible to get a companion on the orbit required to produce the QPO in 1ES\,1927+654, the details of how this is done and what impact the existing AGN disk has on the orbiter is still unclear, thereby motivating further simulations of TDEs and pTDEs interacting with existing accretion disks. 

One of the exciting consequences of this mass-transferring orbiter model is the potential for an observable mHz gravitational waves with the Laser Interferometer Space Antenna (LISA)\cite{Amaro-Seoane2017}. If the QPO is indeed related to a 0.1 $M_\odot$ orbiting companion, then we estimate that the system should produce a LISA detectable signal. As a crude estimate of the LISA detectability of such a source, we treat the system as monochromatic at an orbital frequency of 2.34 mHz, and compute the LISA SNR using the method described in (\citenum{Burdge2020}). Assuming a $1.38\times 10^6\,M_\odot$ BH and a 0.1 $M_\odot$ companion, we find a 4 year LISA SNR of approximately 10 at a distance of 84 Mpc. This is a simplified treatment intended to illustrate that such a source could produce an appreciable signal in LISA. A more careful treatment would involve accounting for the frequency evolution, evolving inclination, and all the constraints imposed by the electromagnetic observables (e.g. knowing the source location, frequency, and frequency evolution as priors). A key takeaway of our analysis is that when compared to Galactic binaries, systems involving a WD+SMBH at similar frequencies are detectable at large distances because the large chirp mass of the system compensates for the distance to such sources.

\subsubsection{Disk Tearing Instabilities \& Precession Models} \label{subsec:instability}

Both the strong dependence of the QPO frequency on the X-ray flux and the increasing RMS with energy suggest a direct connection between the underlying inner accretion flow, the corona, and the QPO. As similar trends are commonly seen in LFQPOs in BHXBs, we explore some of the models used to explain them, despite their frequency differences (see Section \ref{subsec:bhxbs} for more details). A particularly promising class of models used to explain the frequency evolution and hard X-ray dependence is Lense-Thirring precession\cite{Stella1998,Stella1999,Ingram2009,Motta2014a,Motta2014b}. Precession can be invoked through either a single test particle\cite{Stella1998,Stella1999} or a solid body (e.g. the inner hot accretion flow)\cite{Ingram2009}, although recent work has shown that the single particle model is the limiting case of the solid-body model (in the limit of small radial extent)\cite{Motta2018}. A misaligned inner accretion flow with respect to the black hole spin axis, required for precession, could be the result of a TDE hitting the inner disk, which has been proposed to explain the initial outburst of 1ES\,1927+654\cite{Ricci2020,Ricci2021,Masterson2022}. However, the Lense-Thirring precession frequency is rather slow compared to the orbital frequency, except at high spin. To reproduce the observed QPO frequency with Lense-Thirring precession (with a single test particle) would require $a \gtrsim 0.9$ and precession at the ISCO for the highest frequency QPO from March 2024 ($f = 2.34$ mHz). This is not helped by a better treatment of rigid body precession of the inner flow\cite{Franchini2016}, since the frequency can only be increased up to the ISCO frequency. Likewise, it is unclear how such a phenomenon could persist near the ISCO for nearly 2 years. Thus, Lense-Thirring precession is difficult to reconcile with the observed QPO frequencies in 1ES\,1927+654.

A truncated disk and misaligned inner flow can arise from disk tearing instabilities, in which a warped disk breaks due to the differential precession torques exceeding the viscous torques holding the disk together\cite{Ogilvie1999,Dogan2018}. These disk tearing instabilities have been seen in thin disk simulations using both smoothed particle hydrodynamics (SPH)\cite{Nixon2012a,Nixon2012b,Nealon2015,Raj2021} and general relativistic magnetohydrodynamics (GRMHD)\cite{Liska2019,Liska2021,Musoke2023} with different magnetic field configurations. This can lead to discrete rings that can precess on rigid-body Lense-Thirring timescales, but these discrete disks can also modulate the accretion rate in a quasi-periodic fashion. This accretion rate modulation has been suggested as a potential explanation for QPEs, as well as for the heartbeat-like variability in the exotic BHXBs GRS 1915+105 and IGR J17091-3624\cite{Raj2021}. The timescales in these systems are much longer than we observe; to reach the observed timescales in 1ES\,1927+654 would require a significantly smaller break radius. The break radius is expected to depend inversely on the viscosity and scale height\cite{Nixon2012b}, which could explain the increasing frequency with flux (i.e. increasing $\dot{M}$ should also increase $H/R$, thereby shrinking the break radius). However, recent GRMHD simulations with a realistic treatment of magnetic turbulence have shown disagreement with break radius from analytic expectations and SPH simulations\cite{Liska2021,Musoke2023}, and thus, the exact scaling of the break radius is currently poorly understood. It is therefore unclear whether such an instability could explain the QPO in 1ES\,1927+654. 

Additionally, recent GRMHD simulations of a geometrically thin, highly tilted disk have found that in addition to producing LFQPOs with Lense-Thirring precession of the inner accretion flow, HFQPOs ($f \approx 40-55$ Hz for a 10 $M_\odot$ black hole) can also naturally be produced at the radial (or vertical) epicyclic frequency of the truncation radius\cite{Musoke2023,Bollimpalli2024}. This is another promising idea for the QPO in 1ES\,1927+654 -- it allows for a direct connection to the X-ray corona through the inner hot flow and the frequency evolution with flux could be driven by a truncation radius that depends on $\dot{M}$ (although see the above paragraph for complications around determining the truncation radius). However, given the rapid timescales in 1ES\,1927+654, if the QPO is associated with the radial epicyclic frequency, then this requires the disk tearing radius to be $\lesssim 5\, R_g$, the black hole to be rapidly spinning ($a \gtrsim 0.8$), and the mass of the black hole to be on the low end of the best estimate\cite{Li2022}. This is illustrated in Extended Data Figure \ref{fig:radial_epicyclic}, which shows how the mass, spin, and radius are constrained by associating the QPO with the radial epicyclic frequency; only a small part of parameter space is viable for the most rapid QPO in 1ES\,1927+654. However, similar issues are identified for the 450 Hz HFQPO in GRO J1655-40\cite{Strohmayer2001}, suggesting that not all HFQPOs may arise from this mechanism. Lastly, this model also predicts a relatively transient QPO feature, as the accretion time of the inner sub-disk is of order days around a $10^6 \, M_\odot$ SMBH. Thus, the persistence of the QPO in 1ES\,1927+654 for nearly 2 years would require many tearing cycles to persist.

Overall, the disk tearing and precession models provide a promising picture for the QPO in 1ES\,1927+654, as they likely have a strong dependence on the mass accretion rate. However, as with the orbiter model, they run into complications, including the high frequency and long-lived nature of this QPO. Regardless of the exact physics involved in these models, one thing is clear: the QPO must originate from very close to the ISCO.

\subsubsection{Coronal Oscillations} \label{subsec:corona}

The high frequency, increasing strength of the QPO with energy (see Extended Data Figure \ref{fig:rms}), and correlation between the QPO frequency and hard X-ray flux and spectral index (see Extended Data Figure \ref{fig:qpofreq_fluxevol}) may suggest an intrinsic dependence on the compact X-ray corona. Theoretical work has shown that perturbations at the outer boundary of the corona can give rise to magnetoacoustic waves that propagate within the corona\cite{Cabanac2010}. These waves modulate the Comptonization efficiency and drive oscillations on timescales of $f \approx 2\pi c_s / r_c$, where $c_s$ is the sound speed in the plasma and $r_c$ is the radial extent of the corona\cite{Cabanac2010}. For a $T = 2 \times 10^8$ K plasma (from recent NuSTAR observations; S. Laha et al. ApJ, submitted) with a radial extent of 5 $R_g$, this corresponds to $f \approx 0.9$ mHz, comparable to what we find for the QPO in 1ES\,1927+654. The sound speed scales like $T^{1/2}$, so an increasing temperature, decreasing radial extent, or some combination of the two can explain the frequency evolution seen in 1ES\,1927+654. Assuming a constant temperature, the corona would need to shrink in radial extent from roughly 5 $R_g$ (July-August 2022, $f = 0.93$ mHz) to roughly 2 $R_g$ (March 2024, $f = 2.34$ mHz). This is consistent with the typical coronal sizes inferred from reverberation mapping studies of AGN\cite{Kara2016}. This model also predicts that as the radial extent of the corona decreases (and the QPO frequency increases), the X-ray spectrum should soften as the temperature of the seed photons increases. This matches the observed behavior seen in 1ES\,1927+654 (see left panel of Extended Data Figure \ref{fig:qpofreq_fluxevol}) and provides a natural connection between the X-ray spectral shape and the QPO frequency.

A similar connection between the corona and the QPO, inferred from the rapidly evolving QPO frequency, has been suggested in BHXB MAXI J1820\cite{Buisson2019}. However, in this source, the frequency is relatively low ($f \approx 0.1$ Hz, for a stellar mass black hole), implying a significantly larger radial extent of the corona than in 1ES\,1927+654. Additionally, while the corona was found to contract when MAXI J1820 was in the bright, hard state\cite{Kara2019}, the corona actually expanded as the source transitioned from the hard to soft state and launched a transient jet\cite{Axelsson2021,DeMarco2021,Wang2021}. The current spectral state of 1ES\,1927+654 and the recent radio flare\cite{Meyer2024} suggest that the source is more akin to what was seen in the intermediate state with MAXI J1820, where the corona expanded. However, this does not preclude the QPO from being driven by coronal oscillations; as discussed in Section \ref{subsec:bhxbs}, the QPO in 1ES\,1927+654 is not a direct analog to BHXB QPOs, so a direct comparison to the canonical state transition in BHXBs may not be warranted. 

1ES\,1927+654 is one of the only AGN to show the rapid destruction and recreation of the X-ray corona, which lends credence to the idea that this never-before-seen QPO is related to the dynamics and radiative properties of a newly-formed corona. It is prudent, however, to ask why the QPO only appears in the last few years in this model. The top right panel of Figure \ref{fig:overview} highlights that 1ES\,1927+654 is currently quite soft, but not in an unusual spectral state compared to when it was not showing QPOs (e.g. throughout late 2019-2020). In particular, the XMM-Newton observation in November 2020 showed a nearly identical X-ray flux and spectral state, yet no such QPO was detected in this observation. This suggests that there must be another factor, beyond the spectral shape and flux alone, that dictates whether a QPO is present. Interestingly, the corona has also been proposed to be related to the base of a relativistic jet\cite{Markoff2001,Falcke2004,Markoff2005}, and the recent rise in radio flux suggests the potential launching of a newborn jet\cite{Meyer2024}. Thus, the turn-on of the hard X-ray QPO in the corona may be connected to the presence of a relativistic jet. 

\end{methods}

\newpage

\begin{table}
    \centering
    \renewcommand{\tablename}{Extended Data Table}
    \caption{XMM-Newton Observation Details} \label{tab:xmm_obs}
    \begin{tabular}{ccccc}
        \hline
        ObsID & Date & Epoch & Total Exposure (ks) & Cleaned Exposure (ks) \\
        \hline
        \hline
        0902590201 & 2022-07-26 & Jul.-Aug. 2022 & 27.9 & 15.3 \\
        0902590301 & 2022-07-28 & Jul.-Aug. 2022 & 19.0 & 17.1 \\
        0902590401 & 2022-07-30 & Jul.-Aug. 2022 & 19.0 & 17.1 \\
        0902590501 & 2022-08-01 & Jul.-Aug. 2022 & 22.0 & 16.8 \\
        0915390701 & 2023-02-21 & Feb. 2023 & 34.6 & 27.5 \\
        0931791401 & 2023-08-07 & Aug. 2023 & 36.4 & 27.0 \\
        0932392001 & 2024-03-04 & Mar. 2024 & 33.0 & 14.7 \\
        0932392101 & 2024-03-12 & Mar. 2024 & 33.1 & 21.3 \\
        \hline
    \end{tabular}
\end{table}

\begin{figure}
    \centering
    \renewcommand{\figurename}{Extended Data Figure}
    \includegraphics[width=\textwidth]{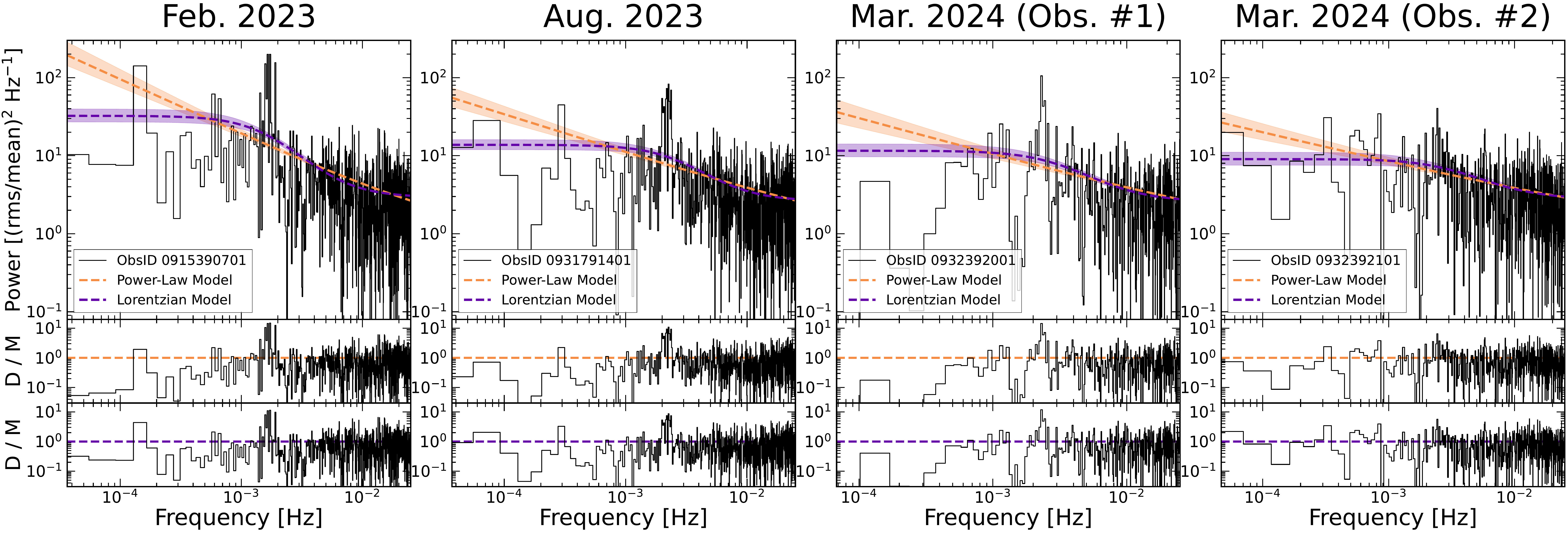}
    \caption{\linespread{1}\normalsize Unbinned PSDs from the 2-10 keV light curves with 20s binning for ObsIDs 0915390701, 0931791401,  0932392001, and 0932392101. The resulting MCMC fit to a power-law or Lorentzian broadband noise model is shown in orange or purple, respectively. The shaded regions show a $1\sigma$ confidence region. The bottom two panels show the data divided by the power-law and Lorentzian models, highlighting the strong QPO feature at roughly 1.7 mHz (ObsIDs 0915390701), 2.2 mHz (ObsID 0931791401), and 2.3 mHz (ObsIDs 0932392001, 0932392101). These broadband noise models were used to simulate power spectra for estimating the statistical significance of the QPOs.}
    \label{fig:broadband}
\end{figure}

\begin{figure}
    \centering
    \renewcommand{\figurename}{Extended Data Figure}
    \includegraphics[width=\textwidth]{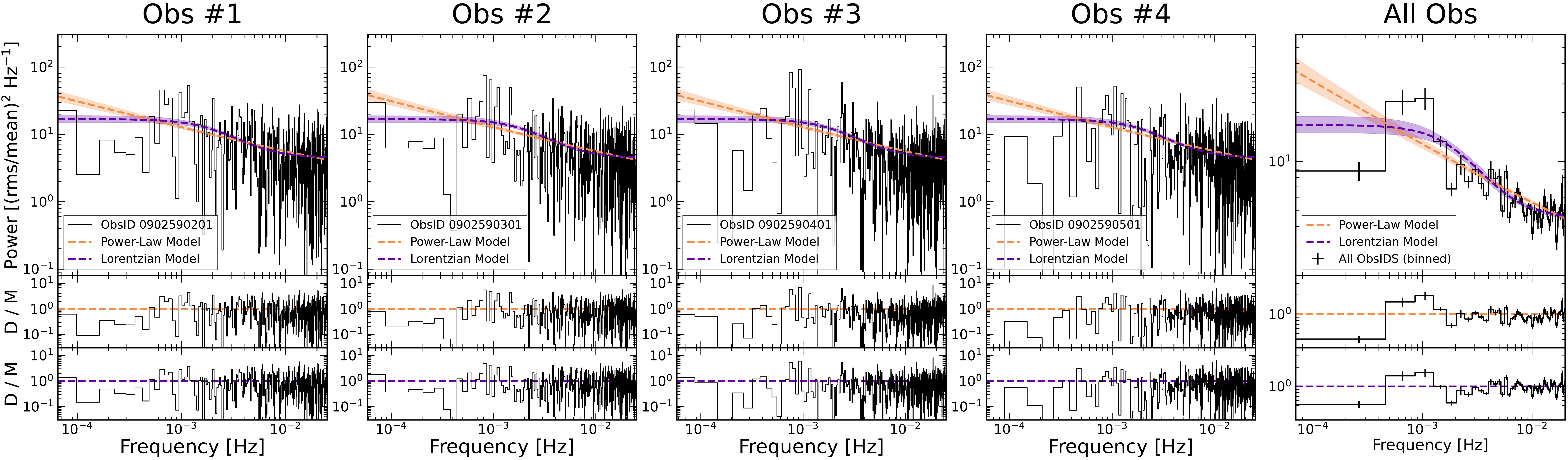}
    \caption{\linespread{1}\normalsize \textit{Left Four Panels:} Unbinned PSDs from the 2-10 keV light curves with 20s binning for the 4 observations taken in July-August 2022 (ObsIDs 0902590201, 0902590301, 0902590401, 0902590501). The resulting MCMC fit to a power-law or Lorentzian broadband noise model is shown in orange or purple, respectively, with the $1\sigma$ confidence intervals shown as shaded regions. This model is fit simultaneously to all of these observations with all parameters tied. The bottom two panels show the data divided by the power-law and Lorentzian models. All four observations show an excess near 0.9 mHz. \textit{Rightmost Panel:} Binned PSD created by averaging the individual PSDs at each frequency and then binning $n = 6$ neighboring frequency bins. The resulting simultaneous fits to the broad band noise are shown again, solely for visual purposes. There is a clear, but broad, excess around 0.9 mHz.}
    \label{fig:aug2022_broadband}
\end{figure}

\begin{ThreePartTable}
\begin{table}  
    \scriptsize
    \centering 
    \renewcommand{\tablename}{Extended Data Table}
    \caption{Details of the QPO in 2-10 keV XMM-Newton Data} \label{tab:qpo}
    \begin{tabular}{ccccccccc}
        \hline
        Epoch & ObsID(s) & $f_\mathrm{QPO}$\tnote{a} & $p_\mathrm{SSE}$\tnote{b} & $p_\mathrm{SSE}$\tnote{c} & $p_\mathrm{AIC}$\tnote{d} & $p_\mathrm{AIC}$\tnote{e} & $Q_\mathrm{QPO}$\tnote{f} & RMS$_\mathrm{QPO}$\tnote{g} \\
         & & (mHz) & (PL) & (Lor) & (PL) & (Lor) &  & \% \\
        \hline
        \hline
        \multicolumn{9}{c@{}}{Significant Detections} \\
        \hline
        Feb. 2023 & 0915390701 & $1.67 \pm 0.04$ & $<10^{-5}$ & $2 \times 10^{-5}$ & $4 \times 10^{-18}$ & $10^{-10}$ & $8.7_{-3.0}^{+5.4}$ & $19.1_{-2.5}^{+3.4}$ \\
        Aug. 2023 & 0931791401 & $2.21 \pm 0.05$ & $1.3 \times 10^{-4}$ & $1.6 \times 10^{-2}$ & $8 \times 10^{-18}$ & $10^{-11}$ & $8.0_{-2.6}^{+4.5}$ & $15.2_{-1.9}^{+2.3}$ \\
        Mar. 2024 & 0932392001 & $2.34 \pm 0.05$ & $5 \times 10^{-5}$ & $1.7 \times 10^{-3}$ & $3 \times 10^{-10}$ & $2 \times 10^{-7}$ & $10.1_{-4.7}^{+13.8}$ & $15.6_{-2.9}^{+4.1}$ \\
        \hline
        \multicolumn{9}{c@{}}{Marginal Detections} \\
        \hline
        Jul.-Aug. 2022 & 0902590201-501 & $0.93 \pm 0.06$ & $\sim 0.5$ & $\sim 0.7$ & $5 \times 10^{-7}$ & $2 \times 10^{-5}$ & $2.3_{-0.5}^{+1.0}$ & $12.1_{-1.5}^{+1.7}$ \\
        Mar. 2024 & 0932392101 & $2.50 \pm 0.18$ & $\sim 0.7$ & $\sim 0.8$ & $0.11$ & $0.16$ & $7.8_{-2.8}^{+12.2}$ & $7.6_{-2.6}^{+2.3}$ \\
        \hline
    \end{tabular}
    \begin{tablenotes}
    \item Uncertainties represent the $1\sigma$ confidence interval
    \item[a] Centroid frequency of the QPO, as measured from fits with an additional Lorentzian for the QPO (with the power-law broadband noise model).
    \item[b] $p$-value from the SSE significance testing, assuming a power-law broadband noise model
    \item[c] $p$-value from the SSE significance testing, assuming a Lorentzian broadband noise model (centroid frequency = 0)
    \item[d] $p$-value from the AIC significance testing, assuming a power-law broadband noise model
    \item[e] $p$-value from the AIC significance testing, assuming a Lorentzian broadband noise model (centroid frequency = 0)
    \item[f] Quality factor of the QPO, defined as $Q = f_\mathrm{QPO}$/FWHM$_\mathrm{QPO}$, as measured from fits with an additional Lorentzian for the QPO (with the power-law broadband noise model). Uncertainties represent the $1\sigma$ confidence interval.
    \item[g] Fractional RMS of the QPO, as measured from the normalization of the additional Lorentzian for the QPO (with the power-law broadband noise model). Uncertainties represent the $1\sigma$ confidence interval.
    \end{tablenotes}
\end{table} 
\end{ThreePartTable}

\begin{figure}
    \centering
    \renewcommand{\figurename}{Extended Data Figure}
    \includegraphics[width=\textwidth]{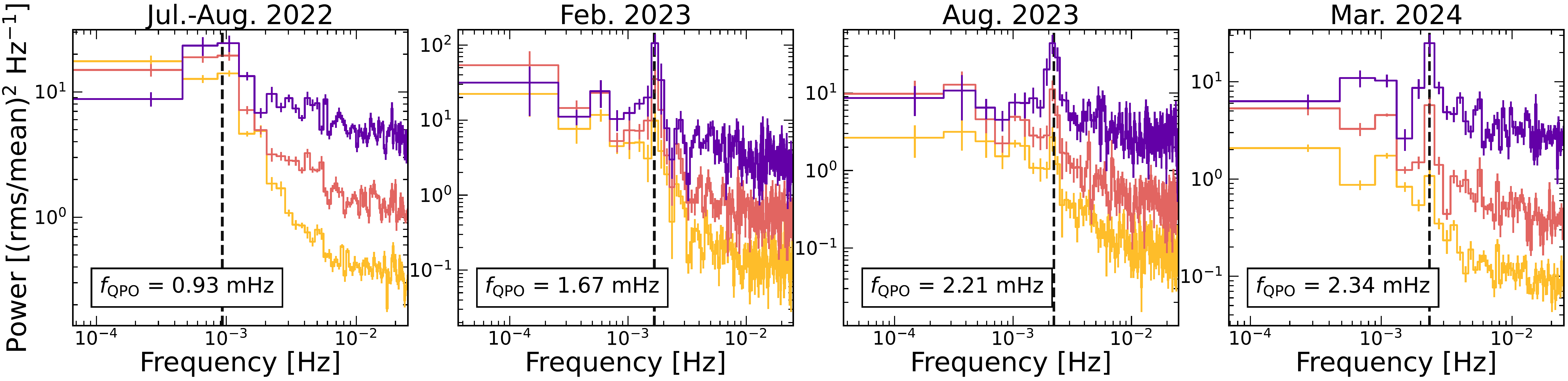}
    \caption{\linespread{1}\normalsize Binned PSDs ($n = 6$ frequencies per bin, standard error uncertainties) for each of the observations showing evidence for a QPO. The yellow, orange, and purple data shows the 0.3-2 keV, 1-4 keV, and 2-10 keV PSDs, respectively. Each column corresponds to a single epoch in time. Note that the July-August 2022 (March 2024) data contains 4 (2) observations taken within roughly 1 week of each other. The lower energy PSDs all show a weaker QPO than the higher energy (2-10 keV) PSDs, and thus, we use the 2-10 keV PSDs for analysis of the QPO.}
    \label{fig:PSD_energy}
\end{figure}

\begin{figure}
    \centering
    \renewcommand{\figurename}{Extended Data Figure}
    \includegraphics[width=\textwidth]{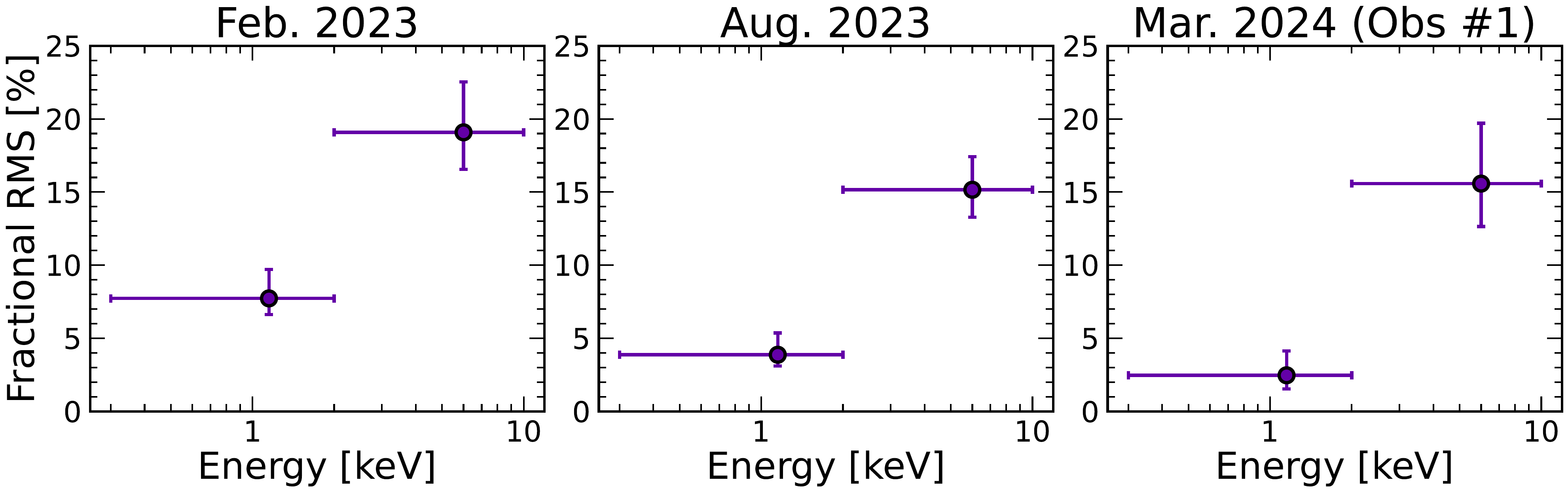}
    \caption{\linespread{1}\normalsize Fractional RMS of the QPO in February 2023, August 2023, and March 2024 (first observation only) in two different energy bands -- 0.3-2 keV and 2-10 keV. The fractional RMS was estimated using the normalization of the best-fitting Lorentzian for the QPO, and the error bars represent 1$\sigma$ confidence intervals from the MCMCs. As in BHXBs, the fraction RMS increases with energy, suggesting that the Comptonized component is what is being modulated on the QPO frequency.}
    \label{fig:rms}
\end{figure}

\begin{figure}
    \centering
    \renewcommand{\figurename}{Extended Data Figure}
    \includegraphics[width=\textwidth]{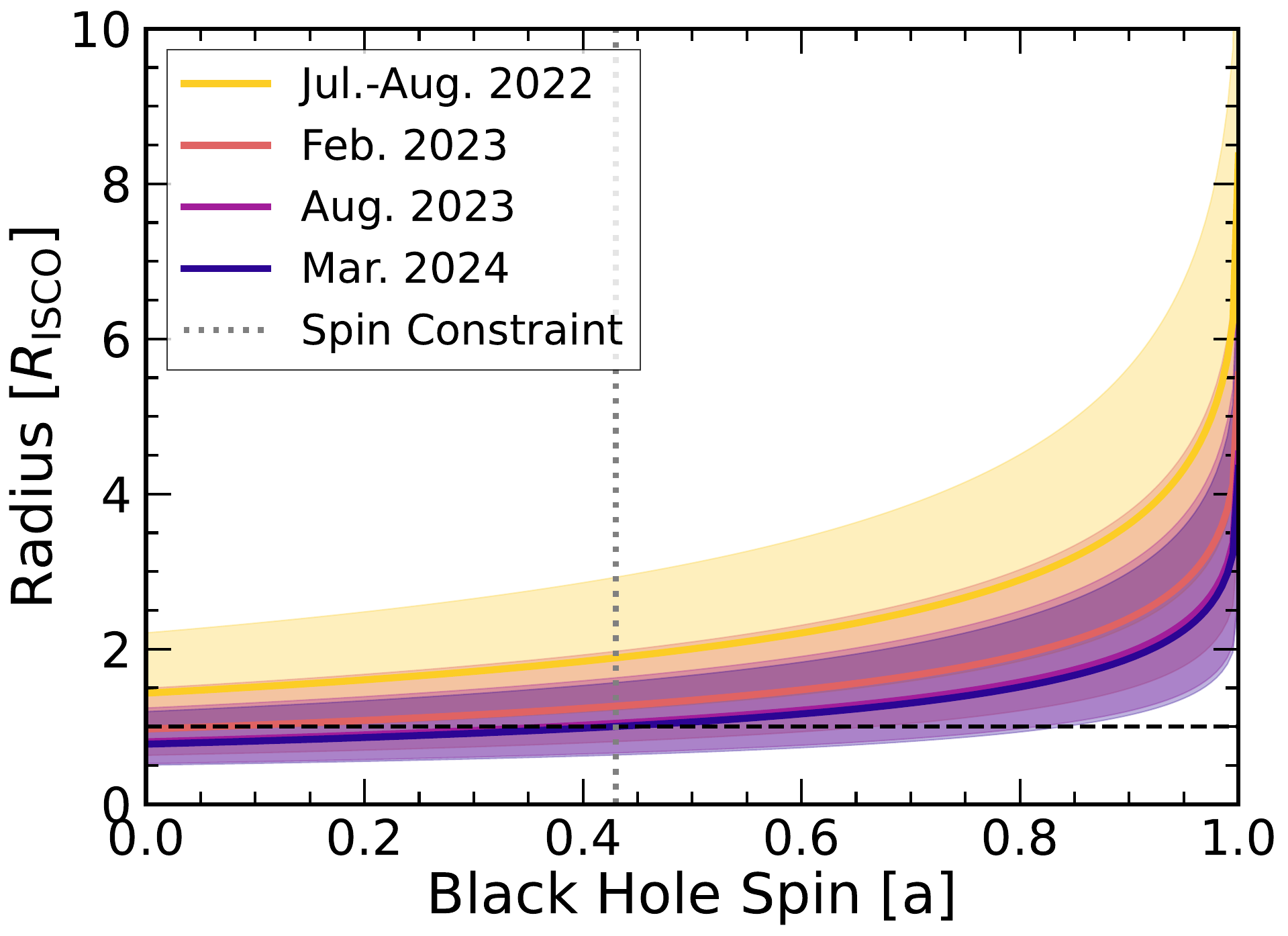}
    \caption{\linespread{1}\normalsize Location of the QPO in terms of the ISCO, assuming that the QPO corresponds to the orbital frequency. The solid lines show the best mass estimate ($1.38 \times 10^6 \, M_\odot$; ref. \citenum{Li2022}), which gives rise to the grey dotted line as a lower limit on the spin of the SMBH. However, the spin constraint is highly sensitive to the mass, which has significant uncertainty. The shaded regions show the effects of this ($1\sigma$) uncertainty\cite{Li2022}.}
    \label{fig:isco}
\end{figure}

\begin{figure}
    \centering
    \renewcommand{\figurename}{Extended Data Figure}
    \includegraphics[width=\textwidth]{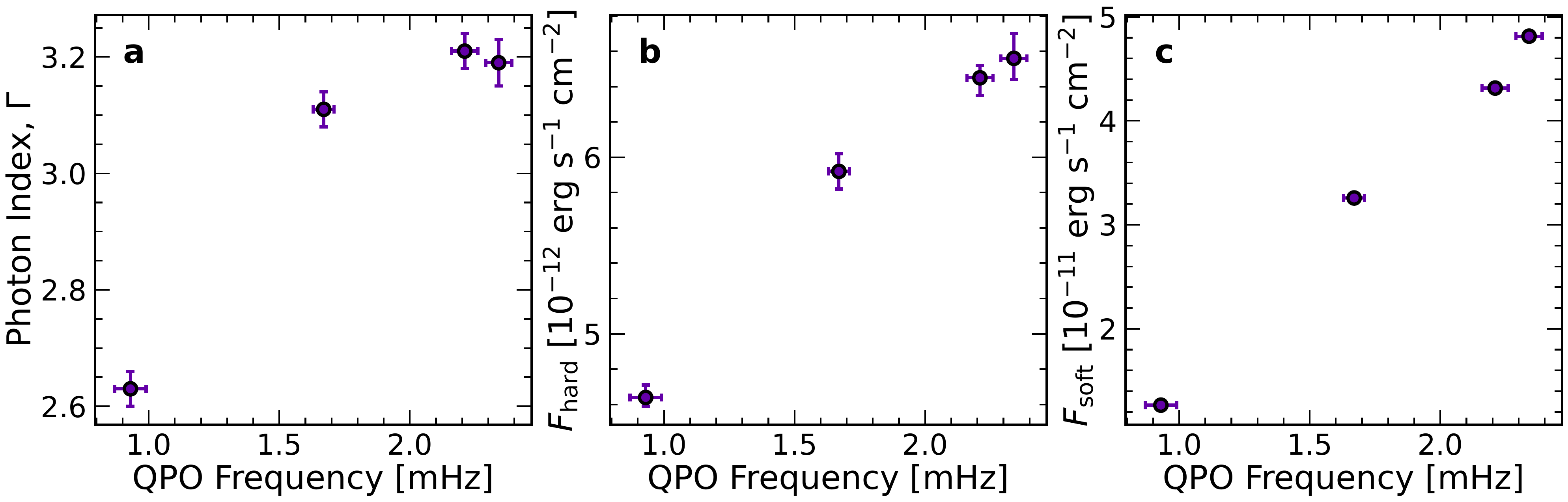}
    \caption{\linespread{1}\normalsize Frequency dependence of the QPO on the spectral shape and X-ray flux, all of which show positive correlations. Error bars represent 1$\sigma$ uncertainty. \textbf{a,} Photon index of the power-law component versus QPO frequency. The photon index was measured by fitting the 0.3-10 keV spectrum with the XSPEC model \texttt{tbabs $\times$ ztbabs $\times$ (zpower + zbbody)}. \textbf{b,} Hard X-ray flux (2-10 keV) versus QPO frequency. \textbf{c,} Soft X-ray flux (0.3-2 keV) versus QPO frequency. }
    \label{fig:qpofreq_fluxevol}
\end{figure}

\begin{figure}
    \centering
    \renewcommand{\figurename}{Extended Data Figure}
    \includegraphics[width=\textwidth]{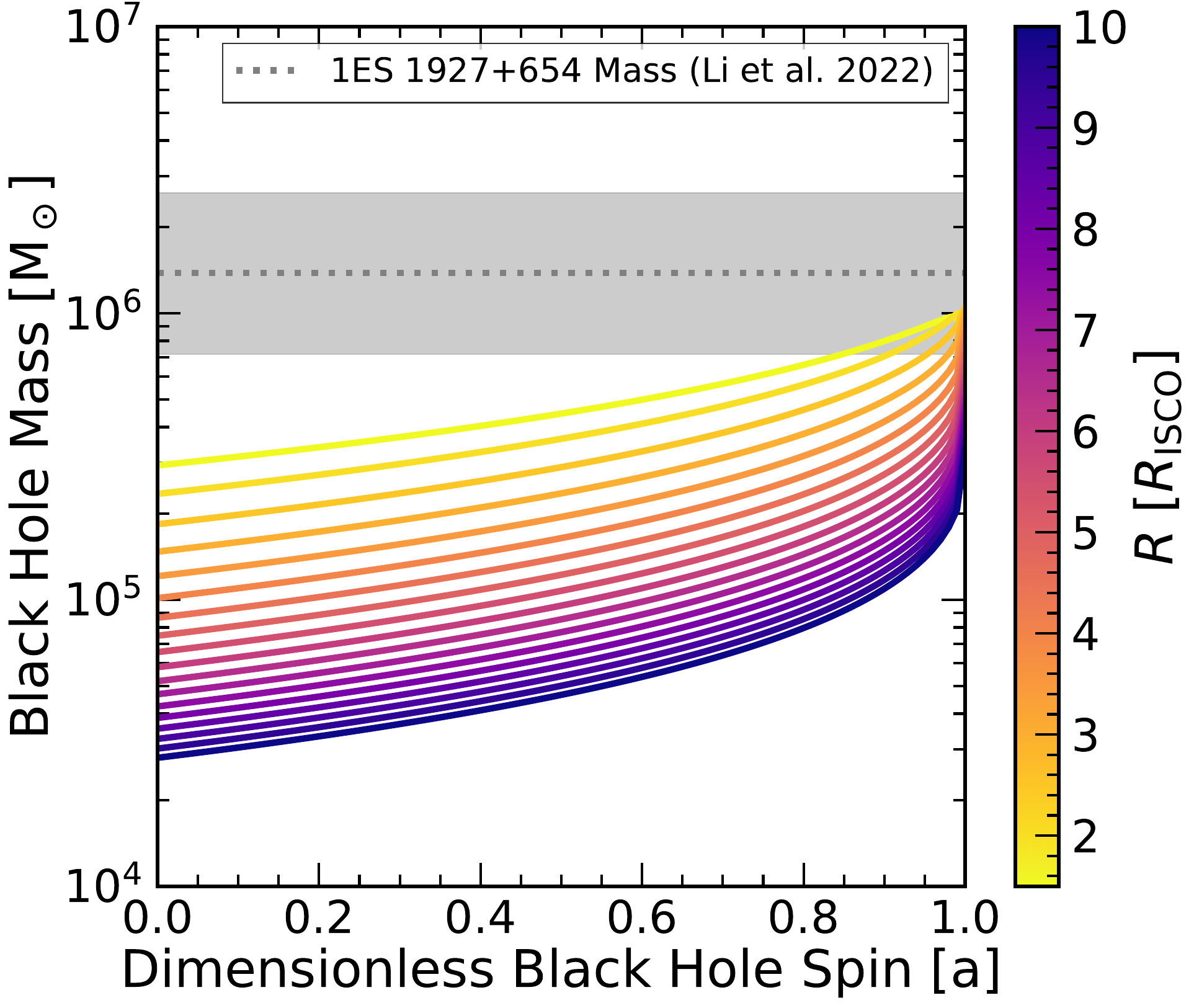}
    \caption{\linespread{1}\normalsize Mass versus spin contours assuming that the most rapid QPO ($f = 2.34$ mHz, March 2024) is the associated with the radial epicyclic frequency at various radii from the black hole. The best mass estimate, from host galaxy scaling relations, is shown in grey, with the shaded regions showing the $1\sigma$ uncertainty region. The QPO can only be associated with the radial epicyclic frequency if the SMBH mass is on the low end of the uncertainty range, the SMBH is rapidly spinning, and the tearing radius is small. Even if the SMBH mass is an order of magnitude lower than the estimate from host galaxy scaling relations, the QPO would still need to be produced within $10\, R_g$ if it is related to the radial epicyclic frequency.}
    \label{fig:radial_epicyclic}
\end{figure}

\clearpage

% \putbib[sup_bib]
% \end{bibunit}
\bibliographystyle{naturemag}
\bibliography{all_bib}

\end{document}